\documentclass[prl,amsmath,amssymb,aps,superscriptaddress,twocolumn,nofootinbib,preprintnumbers
]{revtex4-2}

\usepackage{epsfig}
\usepackage{url}
\usepackage{hyperref}

\usepackage[normalem]{ulem} 

\usepackage{latexsym}
\usepackage{epsfig}
\usepackage{amsmath}
\usepackage{amssymb}
\usepackage{wasysym}
\usepackage{graphicx}
\usepackage{dcolumn}
\usepackage{verbatim}
\usepackage{enumerate,mdwlist}%

\usepackage[titletoc]{appendix}

\usepackage{amsfonts}
\usepackage{fancyvrb} %
\usepackage{tikz} %
\usetikzlibrary{calc}
\usepackage[export]{adjustbox}

\usepackage[normalem]{ulem}%

\usepackage{listings}
\usepackage{xcolor}

\definecolor{comment_red}{rgb}{0.5, 0, 0}
\makeatletter
\g@addto@macro\bfseries{\boldmath}  %
\makeatother

\newcommand{\ba}{\begin{eqnarray}}
\newcommand{\ea}{\end{eqnarray}}

\definecolor{grey}{rgb}{0.4,0.4,0.4}
\definecolor{dullmagenta}{rgb}{0.4,0,0.4}
\definecolor{darkblue}{rgb}{0,0,0.4}
\definecolor{midblue}{rgb}{0,0,0.5}
\definecolor{midred}{rgb}{0.5,0,0}
\definecolor{orange}{rgb}{1,0.5,0}
\definecolor{lightbrown}{rgb}{0.75,0.5,0.25}
\definecolor{tan}{cmyk}{0.14,0.42,0.56,0}
\definecolor{djunglegreen}{cmyk}{0.99,0,0.52,0}
\definecolor{lightgreen}{rgb}{0,1,0}
\definecolor{olivegreen}{cmyk}{0.64,0,0.95,0.40}
\definecolor{midgreen}{rgb}{0.0,0.675,0.0}
\definecolor{darkgreen}{rgb}{0,0.5,0}
\definecolor{ceruleanblue}{rgb}{0.0, 0.2, 0.7}
\definecolor{burgundy}{rgb}{0.5, 0.0, 0.13}
\definecolor{hvred}{RGB}{186,12,47}

\newcommand{\mzc}[1]{}  %
\newcommand{\sxk}[1]{}  %
\hypersetup{
    colorlinks=true,
    linkcolor=ceruleanblue,
    filecolor=ceruleanblue,      
    urlcolor=midblue,
    citecolor=burgundy,
}

\usepackage{booktabs}

\usepackage[all]{xy} %
\usepackage{amsfonts}

\makeatletter
\def\l@subsubsection#1#2{}
\makeatother

\newcommand{\modelname}{\texttt{SVD-stellar-I5-aLIGO}}

\newcommand{\modelclass}{\textsc{ROSD}}

\begin{document}

\title{Effective description of lensed gravitational waves diffracted by stellar fields}

\preprint{IFT-UAM/CSIC-26-124}

\author{Miguel Zumalac\'arregui}
\email{miguel.zumalacarregui@aei.mpg.de}
\affiliation{Max Planck Institute for Gravitational Physics (Albert Einstein Institute) \\
Am Mühlenberg 1, D-14476 Potsdam-Golm, Germany}
\affiliation{Instituto de Física Teórica UAM/CSIC, \\
C/ Nicolás Cabrera 13--15, Universidad Autónoma de Madrid, \\
Cantoblanco, 28049 Madrid, Spain}

\author{Xikai Shan}
\email{xk\_shan@mail.bnu.edu.cn}
\affiliation{Department of Astronomy, Tsinghua University, Beijing 100084, China}

\begin{abstract}
As natural telescopes, Gravitational lenses enable the observation of sources that would otherwise be too distant and faint. Stellar-mass objects, or microlenses, act as impurities in the lens, producing subtle distortions of the source. These effects are necessary to correctly interpret observations, and may in some cases be themselves evidence of gravitational magnification.
Gravitational waves (GWs) observed by ground detectors and magnified by galaxies and clusters will undergo microlensing by fields of stars and remnants: describing these systems requires not only considering a large number of small-scale lenses (microlenses), but also including wave-optics effects, leading to frequency-dependent modulations of the signal. 
Here we present novel models for \textit{Reduced-Order Stochastic Diffraction} (\modelclass), which overcome these challenges in the search for GW lensing signatures: an effective description is synthesized from numerical simulations of wave-optics lensing by stellar fields via a singular value decomposition. 
The procedure yields an optimized orthonormal basis to describe microlensing distortions --- capturing the dominant ``tones'' of the stellar field, or \textit{stellar timbre} --- and a probability density function for the coefficients, which can be used as priors or to verify the consistency with stellar-field lensing.
We present \modelname{} as an example of this model category, discuss the role of truncation order and demonstrate how it can be applied to GW data via injection and recovery in Bayesian parameter estimation.
\modelclass{} can be tailored to account for detector sensitivity and the type of source under analysis, and extended to different microlens populations and external potentials.
\modelclass{} models open a new window to probe small-scale objects (stars, remnants and potentially dark matter) and facilitate the discovery of the most distant compact binary mergers. 

\end{abstract}

\date{\today}

\maketitle

\tableofcontents 

\section{Introduction}

Gravitational lensing, the deflection, magnification, and time delay of signals propagating through gravitational fields, is both a valuable source of information in cosmology and astrophysics, and essential to correctly interpret a wide range of observations~\cite{Schneider:1992,Schneider:2006}. 
In many lensing systems the signal is deflected by a massive object, such as a galaxy or galaxy cluster, with mass $M\sim 10^{11}-10^{15}M_\odot$. 
This macroscopic lens can produce multiple resolved images of the same source. 
Each trajectory, or macroimage, probes a different region of the lens plane and is further affected by compact objects near the line of sight, such as stars, stellar remnants, planets, or compact dark matter candidates. 
These small-scale perturbations are collectively known as microlensing, a phenomenon first developed in the context of stars embedded in a macroscopic lens and later established as a powerful probe of compact objects~\cite{Chang:1979zz,Paczynski:1986,Wambsganss:2006}.

Microlensing plays several distinct roles across astrophysics. 
In Galactic microlensing, a foreground compact object produces a transient magnification of a background star. 
Because the multiple microimages are typically unresolved, the characteristic time-dependent light curve is often the evidence that lensing has occurred. 
This makes microlensing a powerful method to detect faint or dark compact objects, to constrain compact-object contributions to dark matter, and to discover planets around foreground stars~\cite{Paczynski:1986,Mao:2012,Gaudi:2012}. 
In strongly lensed quasars, microlensing by stellar fields in the lensing galaxy modifies the relative magnification of the macroimages, produces long-term variability, and induces chromatic magnification because different emission regions have different angular sizes~\cite{Chang:1979zz,Schechter:2002dm,Vernardos:2023hbu}. 
These effects are a nuisance for some applications, such as measuring macroimage flux ratios, but they also provide a rare probe of quasar accretion discs, broad-line regions, and the stellar mass distribution in the lens.

Microlensing is also important for transient and highly magnified sources. 
For strongly lensed supernovae, microlensing can alter the observed light curves and hence affect time-delay measurements, while also carrying information about the expanding photosphere and progenitor system~\cite{Suyu:2023yvy}. 
Near the caustics of galaxy clusters, stars and stellar remnants in the lens corrugate the macroscopic caustic into a network of microcaustics. 
This enables the detection of individual stars at cosmological distances and makes caustic-crossing events sensitive probes of compact objects and dark matter substructure~\cite{Venumadhav:2017pps,Oguri:2017ock,Diego:2018fzr,Weisenbach:2024}. 

Gravitational waves (GWs) provide a qualitatively different microlensing regime. 
Lensing is an inevitable propagation effect for GWs, but its observational imprint depends on the relation between the wavelength and the characteristic scale of the lens. 
For electromagnetic observations, stellar microlensing is usually well described by geometrical optics. 
For GWs in the ground-based detector band, however, the wavelength can be comparable to the gravitational radius of stellar remnants or intermediate-mass compact objects, corresponding to dimensionless frequencies $GM_{Lz} \sim \mathcal{O}(1)$ for stellar-mass lenses~\cite{Takahashi:2003ix,Christian:2018vsi,Dai:2018enj}. 
Microlensing can then imprint frequency-dependent diffraction and interference patterns on the waveform, rather than merely changing its overall magnification. 
This was first explored for simple lens models and isolated compact objects~\cite{Takahashi:2003ix,Dai:2018enj,Cheung:2020okf,Yeung:2021roe,Yeung:2021chy}, and later extended to microlenses embedded in macroscopic lens potentials and realistic stellar populations~\cite{Diego:2019lcd,Diego:2019rzc,Oguri:2020ldf,Mishra:2021xzz,Meena:2022unp,Oguri:2022zpn,Shan:2022xfx,Shan:2023ngi,Shan:2023qvd,Smith:2025axx,Meena:2025gry,Shan:2025jpt}. 
These wave-optics signatures may help identify lensed GWs, infer properties of the lensing environment, and distinguish microlensing from other waveform distortions.

Computing these signatures is itself challenging. 
The amplification factor is given by a highly oscillatory diffraction integral over the lens plane, with pathological convergence properties. 
Early studies of wave-optics microlensing already recognized the difficulty of evaluating diffraction integrals~\cite{Ulmer:1994ij,Jaroszynski:1995cd}. 
Recent work has produced substantial progress in evaluating this integral, including a variety of methods and public software packages for wave-optics lensing~\cite{Villarrubia-Rojo:2024xcj,Feldbrugge:2019fjs,Tambalo:2022plm,Shan:2022xfx,Shan:2024min,Cheung:2024ugg,Yeung:2024pir,Ephremidze:2026era}. 
These developments have made it possible to compute accurate amplification factors for increasingly realistic lens models. 

However, computation alone does not solve the inverse problem: the information encoded in a diffracted waveform must be extracted in order to be interpreted astrophysically.
Given the difficulties in wave-optics computations, most current GW analyses adopt simplified deterministic models, such as isolated symmetric lenses~\cite{Cheung:2024ugg,Wright_2022,Caldarola:2025oxr,lvklensingO3a,lvklensingO3b,lvklensingO4a} or point lenses embedded in an external potential~\cite{Goyal:2025eqo,Shan:2025dcd}, neglect wave-optics phenomena altogether~\cite{Liu:2023ikc,Chan:2025pdf,Hu:2025lhv} or search for unmodeled correlated residuals between detectors, without the structure of lensing-induced distortions~\cite{Chakraborty:2024mbr,Seo:2025dto,Chakraborty:2025maj}. 
Inference in GWs diffracted by realistic and complex matter distributions remains an open problem.

The central difficulty for realistic microlensing is the high dimensionality and stochastic nature of the lensing field. 
A macroimage in a galaxy-scale lens is affected not by a single object, but by a population of stars and remnants embedded in the external gravitational potential of the macrolens. 
The resulting diffraction pattern depends not only on the external shear, but also on the positions and masses of a vast number of individual objects. 
Directly sampling over all these degrees of freedom in GW parameter estimation is therefore impractical. 
Advances have been made using simulation-based inference~\cite{Su:2025xry}, but these methods require expensive model training, and their results are difficult to interpret. 
An outstanding task is to formulate a description of microlensing that is general enough to account for the wide range of possible matter distributions, detailed enough to encode the specific predictions of stellar fields, and simple enough to be applicable to real GW data.

In this work we develop an effective stochastic description of microlensing diffraction by constructing a low-dimensional basis from a large suite of wave-optics simulations. 
This strategy is inspired by reduced-order and surrogate models used in GW waveform modelling~\cite{Field:2013cfa,Canizares:2014fya,Blackman:2015pia,Deka:2025vzx}, and by the broader use of reduced or regularized representations in gravitational-lens inverse problems~\cite{Warren:2003na,Suyu:2006fd}. 
We refer to this class of effective description as \textit{Reduced-Order Stochastic Diffraction} (\modelclass{}) models. 
The implementation developed in this work is denoted \modelname: 
\texttt{SVD} specifies the compression method, \texttt{stellar} the microlens population, \texttt{I5} the macropotential (image type and maximum magnification), and \texttt{aLIGO} the detector-noise weight used in the inner product.
This provides a compact phenomenological model that can be used in GW inference while retaining a connection to the underlying microlens population.
The basis modes represent the characteristic distortions that a stellar field imprints on the waveform: in analogy with how an acoustic environment --- a concert hall, a small room, the open air --- colours the sound of an instrument (here, the GW source), the lensing environment modulates the emitted signal. We refer to this signature as the \textit{stellar timbre}, in parallel with the ``dark timbre'' imprinted by dark-matter structures~\cite{Urrutia:2024pos,Choi:2026lsa}.

The paper is outlined as follows:
Sec.~\ref{sec:lensing} recaps the formalism of wave-optics lensing and the properties of stellar fields. 
Sec.~\ref{sec:modelbuild} presents the construction of an effective model via singular value decomposition (SVD).
Sec.~\ref{sec:properties} discusses the properties of the model in connection to astrophysical parameters.
Sec.~\ref{sec:performance} analyzes the performance of the effective model in terms of truncation, parameter estimation, and the possibility to connect to other lens models. 
We discuss our results and future prospects in Sec.~\ref{sec:conclusions}.

\section{Diffraction by stellar fields}
\label{sec:lensing}

Let us first introduce the basic equations and concepts involved in wave-optics lensing 
(Sec.~\ref{sec:lensing_wo}) before describing the stellar microlensing fields used in this work 
(Sec.~\ref{sec:lensing_stellar}). 
We work in natural units with $G=c=1$.

\subsection{Wave-optics lensing} \label{sec:lensing_wo}

Following Ref.~\cite{Villarrubia-Rojo:2024xcj}, the effect of lensing in the frequency domain is encoded in the \textit{amplification factor}
\begin{equation}\label{eq:F_def}
F(f) \equiv \frac{\tilde h(f)}{\tilde h_0(f)} ,
\end{equation}
where $\tilde h(f)$ is the lensed waveform and $\tilde h_0(f)$ is the waveform that would be observed in the absence of lensing. 
In the thin-lens approximation, wave-optics lensing is described by the diffraction integral
\begin{equation}
F(w,\boldsymbol{y},\boldsymbol{\theta})
=
\frac{w}{2\pi i}
\int d^2\boldsymbol{x}\;
\exp\!\left[i\,w\,\phi(\boldsymbol{x},\boldsymbol{y};\boldsymbol{\theta})\right] .
\label{eq:F_waveoptics}
\end{equation}
Here $\boldsymbol{x}$ and $\boldsymbol{y}$ are dimensionless coordinates in the lens and source planes, respectively, and $\boldsymbol{\theta}$ denotes the parameters of the lens model. 
For axisymmetric lenses, $y=|\boldsymbol{y}|$ is the usual impact parameter in units of the chosen lens-plane scale. 
The corresponding physical coordinates are
\begin{equation}
    \boldsymbol{x}=\frac{\boldsymbol{\xi}}{\xi_0},
    \qquad
    \boldsymbol{y}=\frac{\boldsymbol{\eta}}{\eta_0},
    \qquad
    \eta_0=\frac{D_S}{D_L}\xi_0 ,
\end{equation}
where $\xi_0$ is an arbitrary length scale in the lens plane, while $D_L$, $D_S$, and $D_{LS}$ are angular-diameter distances to the lens, to the source, and between lens and source.

The dimensionless frequency is
\begin{equation}
    w=8\pi M_{Lz} f ,
    \label{eq:w_def}
\end{equation}
where $M_{Lz}=(1+z_L)M_L$ is the redshifted lensing mass scale. 
The mass scale $M_L$ is fixed by the choice of $\xi_0$,
\begin{equation}
    M_L
    =
    \frac{\xi_0^2}{4}
    \frac{D_S}{D_LD_{LS}} .
    \label{eq:ML_xi0}
\end{equation}
Equivalently, the physical time delay is
\begin{equation}
    t_d(\boldsymbol{x},\boldsymbol{y})
    =
    4M_{Lz}\,
    \phi(\boldsymbol{x},\boldsymbol{y}) .
    \label{eq:td_phi}
\end{equation}

The dimensionless Fermat potential is
\begin{equation}
\phi(\boldsymbol{x},\boldsymbol{y};\boldsymbol{\theta})
=
\frac{1}{2}\,|\boldsymbol{x}-\boldsymbol{y}|^2
-\psi(\boldsymbol{x};\boldsymbol{\theta})
+\phi_0(\boldsymbol{y}) .
\label{eq:FermatPhi}
\end{equation}
The first term is the geometrical path-length difference, while $\psi$ is the projected lensing potential. 
The additive term $\phi_0$ fixes an arbitrary overall phase convention and can be chosen for convenience. 
For an isolated GW signal, changes in this constant are degenerate with the coalescence phase and time.

The lensing potential is related to the projected surface density by
\begin{equation}
    \nabla^2 \psi(\boldsymbol{x}) = 2\kappa(\boldsymbol{x}),
    \qquad
    \kappa(\boldsymbol{x}) \equiv 
    \frac{\Sigma(\xi_0\boldsymbol{x})}{\Sigma_{\rm crit}},
    \label{eq:kappa_def}
\end{equation}
where
\begin{equation}
    \Sigma_{\rm crit}
    =
    \frac{1}{4\pi}
    \frac{D_S}{D_LD_{LS}}
\end{equation}
in units $G=c=1$. 
The geometrical-optics image positions are the stationary points of the Fermat potential,
\begin{equation}
    \nabla_{\boldsymbol{x}}\phi = 0,
    \qquad
    \boldsymbol{y}
    =
    \boldsymbol{x}
    -
    \boldsymbol{\alpha}(\boldsymbol{x}),
    \qquad
    \boldsymbol{\alpha}=\nabla\psi .
    \label{eq:lens_equation}
\end{equation}
The signed magnification of each image is
\begin{equation}
    \mu_j
    =
    \left[
    \det\left(
    \frac{\partial \boldsymbol{y}}{\partial \boldsymbol{x}}
    \right)_{\boldsymbol{x}_j}
    \right]^{-1}.
    \label{eq:mu_def}
\end{equation}

In the high-frequency limit, $w\gg1$, the diffraction integral can be evaluated by stationary phase. 
The amplification factor then becomes a coherent sum over geometrical-optics images,
\begin{equation}
    F(f)
    \simeq
    \sum_j
    |\mu_j|^{1/2}
    \exp\!\left[
        2\pi i f t_{d,j}
        - i\pi n_j
    \right],
    \label{eq:F_geo}
\end{equation}
where $t_{d,j}$ is the time delay of image $j$ and $n_j$ is the corresponding Morse phase. 
Wave-optics effects become important when $w\lesssim {\cal O}(1)$, or when the time delays between unresolved microimages are comparable to the inverse GW frequency. 
In this regime, lensing cannot be described only by magnifications and time delays; the full frequency-dependent amplification factor must be computed.

Many methods to solve the diffraction integral \eqref{eq:F_waveoptics} have been developed~\cite{Ulmer:1994ij,Diego:2019lcd,Feldbrugge:2019fjs,Mishra:2021xzz,Tambalo:2022plm,Cheung:2024ugg,Villarrubia-Rojo:2024xcj,Yeung:2024pir}. 
These methods enable accurate calculations for isolated lenses and smooth lens models, but realistic stellar fields pose an additional difficulty: the lens plane contains many compact objects, producing a complicated Fermat surface with many microimages and caustics. 
Brute-force integration is therefore inefficient, while purely geometrical methods miss the diffraction and interference effects relevant for GWs.

We evaluate the diffraction integral using the adaptive sampling code developed in Refs.~\cite{Shan:2022xfx,Shan:2024min}.%
\footnote{\url{https://github.com/xkshan97/Microlensing_Wave_Effect}}
The code is designed to compute wave-optics amplification factors for stellar microlensing fields embedded in an external macrolens. 
Rather than sampling the lens plane uniformly, it combines the tree algorithm introduced in Refs.~\cite{Zheng:2022vfq, Chen:2021ftm} with an adaptive refinement algorithm to resolve the regions of the Fermat surface that contribute most strongly to the diffraction integral, including neighborhoods of microimages, microlensing points, and caustic structures.
This makes it possible to generate large ensembles of amplification factors for stochastic microlens populations, which form the input data for the reduced-order description developed below.

\subsection{Stellar fields in an external potential} \label{sec:lensing_stellar}

We now specialize to the local lensing region around a macroimage produced by a galaxy, group, or cluster. 
The lensing potential can be decomposed into a smooth external contribution and a population of compact microlenses,
\begin{equation}\label{eq:macro_micro_potential}
\psi(\boldsymbol{x},\boldsymbol{\theta})
=
\psi_{\rm ext}(\boldsymbol{x})
+
\sum_j \psi_j(\boldsymbol{x}) 
+ \phi_{-}(\boldsymbol{x}).
\end{equation}
The external potential describes the total lens model expanded locally around the macroimage, while the summation accounts for stars and stellar remnants near the line of sight, and $\phi_{-}(\boldsymbol{x})$ is the negative mass sheet introduced to compensate for the discrete object contribution.

For the statistically homogeneous stellar fields considered in this work, the source position can be fixed without loss of generality. 
We therefore set $\boldsymbol{y}=0$ in the following. 
This simplification should not be used for inhomogeneous situations, such as sources near macroscopic caustics, where the impact parameter controls the local microlensing statistics~\cite{Venumadhav:2017pps,Oguri:2017ock,Diego:2018fzr}.

The external potential is taken to be quadratic,
\begin{equation}
\psi_{\rm ext}(\boldsymbol{x},\kappa,\gamma)
=
\frac{\kappa}{2}|\boldsymbol{x}|^2
+
\frac{\gamma}{2}
\left(x_1^2-x_2^2\right) ,
\label{eq:psi_external}
\end{equation}
where $\kappa$ is the total convergence (the sum of the smooth component $\kappa_{\rm s}$ and the stellar component $\kappa_\star$), and $\gamma$ is the external shear.
The lens-plane coordinates have been rotated so that the shear tensor is diagonal. 
The corresponding macromodel magnification is
\begin{equation}\label{eq:mu_macro}
    \mu_{\rm macro}
    =
    \left[
    (1-\kappa)^2-\gamma^2
    \right]^{-1}.
\end{equation}
This is the magnification of the macromodel obtained after averaging over the microlens population. 

Throughout this work we assume the following relation between convergence and shear
\begin{equation} \label{eq:sie_macro_kappa}
    \kappa = \gamma \,,
\end{equation}
This relation holds for the SIS and, more generally, for the singular isothermal ellipsoid (SIE) mass distribution, often used to describe galaxy-scale lenses~\cite{Kormann:1994}. 
For a local minimum (Type-I), this gives
\begin{equation}\label{eq:sie_macro_mu}
    \mu_{\rm macro}   = \frac{1}{1-2\gamma} \,.
\end{equation}
We note that wave-optics lensing predictions are identical for all gravitational potentials related by a mass-sheet transformation, which is explicitly broken by the macro-potential choice \eqref{eq:sie_macro_kappa}, see Ref.~\cite{Goyal:2025eqo}.

The potential of an individual point mass is
\begin{equation}
    \psi_j(\boldsymbol{x})
    =
    \tilde m_j
    \ln
    \left|
    \boldsymbol{x}-\boldsymbol{x}^{m}_{j}
    \right| ,
    \label{eq:psi_point_lens}
\end{equation}
where $\boldsymbol{x}^{m}_{j}$ is the position of the microlens and
\begin{equation}
    \tilde m_j \equiv \frac{M_j}{M_L}
\end{equation}
is its mass in units of the lensing mass scale defined in Eq.~\eqref{eq:ML_xi0}. 
With the convention $\nabla^2\psi=2\kappa$, the point masses contribute
\begin{equation}
    \kappa_\star(\boldsymbol{x})
    =
    \pi
    \sum_j
    \tilde m_j\,
    \delta^{(2)}
    \left(
    \boldsymbol{x}-\boldsymbol{x}^{m}_{j}
    \right) ,
\end{equation}
so that, over a region of area $A$, the mean stellar convergence is
\begin{equation}
    \kappa_\star
    =
    \frac{\pi}{A}
    \sum_{j\in A}
    \tilde m_j .
    \label{eq:kappastar_mean}
\end{equation}

The negative mass sheet used in the simulation follows Ref.~\cite{Zheng:2022vfq}, where it is defined as
\begin{equation}
    \phi_{-}(\boldsymbol{x})
    =
    -\frac{\kappa_\star}{2\pi}
    \int_A
    \ln
    \left|
    \boldsymbol{x}
    -
    \boldsymbol{x}'
    \right|^2
    \,d^2\boldsymbol{x}' .
\end{equation}
Here, $A$ is the area of the microlensing stellar field set in the simulation.
The full Fermat potential for one stellar-field realization is therefore
\begin{equation}
\begin{aligned}
\phi(\boldsymbol{x})
=&\,
\frac{1}{2}|\boldsymbol{x}|^2
-
\frac{\kappa}{2}|\boldsymbol{x}|^2
-
\frac{\gamma}{2}
\left(x_1^2-x_2^2\right)  \\
&-
\sum_j
\tilde m_j
\ln
\left|
\boldsymbol{x}-\boldsymbol{x}^{m}_{j}
\right|
-
\phi_{-}(\boldsymbol{x}) .
\end{aligned}
\label{eq:stellar_field_fermat}
\end{equation}

\begin{figure}
    \centering
    \includegraphics[width=0.99\linewidth]{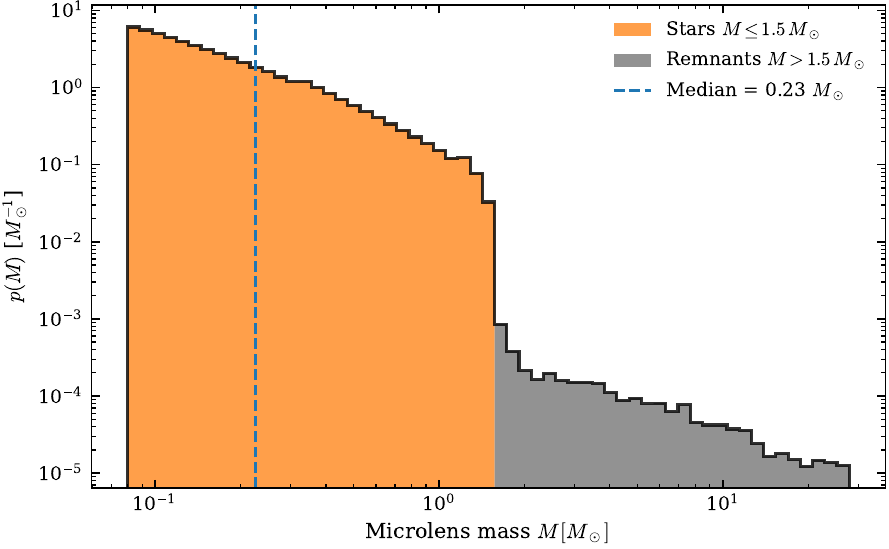}
    \caption{Mass distribution of microlenses in the simulation ensemble.
    The histogram  shows the normalized probability density $p(M)$ of microlens masses drawn from 100 realizations. Stellar lenses and compact remnants are shown in different colors, but share a common normalization.}
    \label{fig:mass_function}
\end{figure}

The microlens population is generated by drawing compact objects from a stellar and remnant mass function and distributing them randomly in the lens plane with mean convergence $\kappa_\star$. 
The stellar component follows the Chabrier initial mass function (IMF)~\cite{Chabrier:2003ki}, with
\begin{equation}
    M_j \in [0.1,1.5]\,M_\odot ,
\end{equation}
while the remnant component follows the initial--final mass relation of Ref.~\cite{spera_remnant_mass_function}, as described in Ref.~\cite{Shan:2024min}.
The mass distribution employed is shown in Fig.~\ref{fig:mass_function}.

For the realizations used here, the remnant component is assigned $20\%$ of the total mass in living stars ($16.7\%$ of the total microlens mass), with maximum remnant masses of approximately $27\,M_\odot$.
This value is motivated by expectations from standard stellar-population models.
For a Chabrier IMF, Madau and Dickinson~\cite{Madau:2014bja} find a dark-remnant mass fraction $D=0.19$ per stellar generation and a return fraction $R=0.41$.
Expressed relative to the mass retained in long-lived stars and remnants, this corresponds to $\frac{D}{1-R}\simeq 0.32 $. Thus a $20\%$ remnant fraction is a plausible, and not especially aggressive, choice for an evolved stellar population.
For comparison, Ref.~\cite{Meena:2022unp} adopted a remnant-to-stellar mass fraction of $10\%$, while noting that stellar-population models allow values in the range $\sim 5$--$20\%$.

The lensing configurations are sampled over external shear, stellar convergence, and lens/source redshifts. Specifically, we first draw the lens and source redshifts independently from uniform proposal distributions,
\begin{equation}
    z_L \sim \mathcal{U}(0.1,\,2.0),
    \qquad
    z_S \sim \mathcal{U}(0.15,\,2.05),
\end{equation}
and retain only configurations satisfying $z_S>z_L$. We further impose $\gamma=\kappa$, and independently propose
\begin{equation}
    \kappa \sim \mathcal{U}(0.1,\,0.4),
    \qquad
    \kappa_{\rm stellar} \sim \mathcal{U}(0.1,\,0.4),
\end{equation}
where $\kappa_{\rm stellar}$ is the convergence in living stars, retaining only samples satisfying
\begin{equation}
    \kappa_\star = 1.2\,\kappa_{\rm stellar} \leq \kappa .
\end{equation}
Here $\kappa_\star$ is the total convergence in compact objects, Eq.~\eqref{eq:kappastar_mean}, which includes the remnant component contributing an additional $20\%$ of the mass in living stars.
The condition therefore ensures that compact objects do not over-saturate the total convergence at the image position, i.e.~that the smooth component $\kappa_{\rm s}=\kappa-\kappa_\star$ remains non-negative.
Thus, $\kappa=\gamma$ is imposed for the accepted lens configuration, while $\kappa_{\rm stellar}$ is drawn jointly with $\kappa$ through the above rejection-sampling condition.
Given Eq.~\eqref{eq:sie_macro_mu}, these choices limit the range of external magnification to $\mu_{\rm macro}\leq 5$.

The distributions above are proposal distributions, chosen to cover the parameter space of lensing configurations evenly, rather than astrophysically realistic ones.
In particular, the joint redshift distribution is not weighted by the strong-lensing probability: configurations with $z_L$ close to $z_S$ correspond to a large critical surface density $\Sigma_{\rm cr}$, and are therefore rare, while for a given source the most probable lens redshift lies roughly halfway between observer and source.
As discussed in Sec.~\ref{sec:properties_macropars}, at fixed $\kappa$, $\gamma$, and $\kappa_\star$ the redshifts affect the amplification factor only through the redshifted microlens masses, $\propto(1+z_L)$, so the basis construction requires adequate coverage of $z_L$ rather than a realistic joint redshift distribution.
Astrophysically weighted ensembles, e.g.~for population studies, can be obtained by reweighting the realizations (Sec.~\ref{sec:properties_priors}) or through dedicated simulations; we leave this to future work.

\section{Model Construction} \label{sec:modelbuild}

\begin{figure*}
    \centering
    \includegraphics[width=\linewidth]{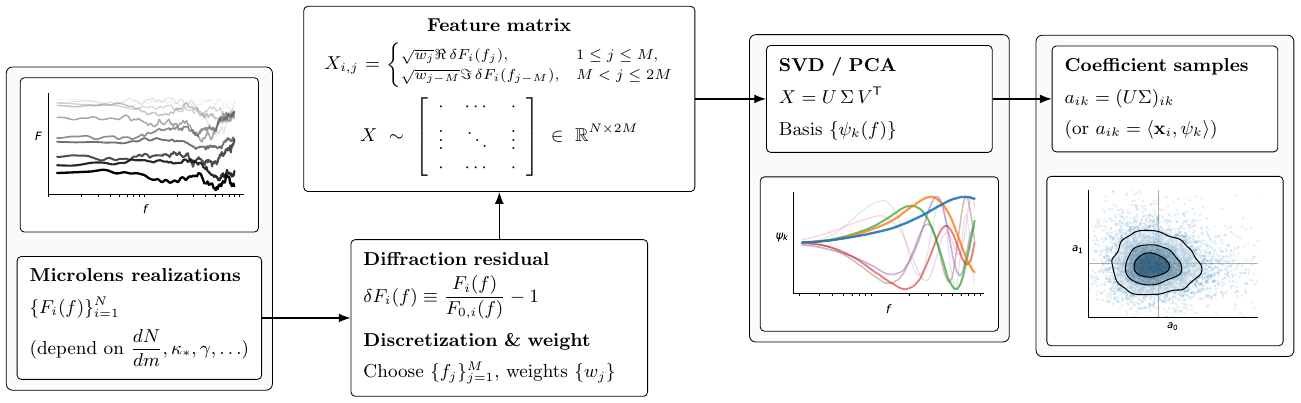}
    \caption{Outline of the microlensing-basis construction. 
    Amplification factors from microlensing realizations are converted into diffraction residuals by factoring out the smooth macroscopic contribution (Sec.~\ref{sec:model_residual}). 
    A frequency-dependent weight is then introduced to reflect detector sensitivity (Sec.~\ref{sec:model_innerproduct}), and the weighted residuals are discretized into a feature matrix. 
    The singular value decomposition of this matrix provides an orthonormal basis ordered by decreasing singular value (Sec.~\ref{sec:model_svd}). 
    The same procedure also yields the distribution of basis coefficients, which can be used as a prior in parameter estimation and to study correlations with macroscopic lens parameters.}
    \label{fig:master_diagram}
\end{figure*}

We now describe the construction of a \textit{Reduced-Order Stochastic Diffraction} (\modelclass{}) model. 
The construction below defines the \modelname{} model:
The labels specify the compression method, microlens population, macroimage properties (type and maximum macromagnification), and inner-product weight. 
Other choices of these ingredients will define different members of the \modelclass{} family.
We first define the microlensing residual (Sec.~\ref{sec:model_residual}) and the inner product used to compare residuals in frequency space (Sec.~\ref{sec:model_innerproduct}). 
We then use a singular value decomposition to construct the microlensing basis and its coefficients (Sec.~\ref{sec:model_svd}). 
The procedure is summarized in Fig.~\ref{fig:master_diagram}.

\subsection{Wave-optics residual} \label{sec:model_residual}

The amplification factor $F(f)$, defined in Eq.~\eqref{eq:F_def}, contains both the smooth macroscopic lensing contribution and the frequency-dependent diffraction produced by the microlens population. 
The former includes an overall magnification, phase, and arrival-time shift. 
For a single GW image these quantities are largely degenerate with the luminosity distance, coalescence phase, and merger time, and should therefore be removed before constructing a basis for microlensing diffraction.

We define the smooth macroscopic contribution as
\begin{equation}\label{eq:F_macro}
F_0(f) \equiv 
\sqrt{|\mu_{\rm macro}|}\,
e^{i(\varphi_0 + 2\pi f\, t_0)} ,
\end{equation}
where $\mu_{\rm macro}$ is the macroscopic magnification defined in Eq.~\eqref{eq:mu_macro}, while $\varphi_0$ and $t_0$ are a constant phase and time shift. 
For the type-I macroimages considered in this work, the macroscopic Morse phase can be absorbed into $\varphi_0$. 
For type-II macroimages, the additional phase shift can become observable in the presence of higher harmonics and should be reintroduced in waveform-level applications~\cite{Dai:2017huk,Ezquiaga:2020gdt}. We will consider other macro-image types in future work.

We define the complex microlensing residual
\begin{equation}\label{eq:delta_F}
\delta F(f) \equiv \frac{F(f)}{F_0(f)} - 1 .
\end{equation}
This residual isolates the frequency-dependent part of the amplification factor associated with stellar microlensing. 
By construction, $\delta F(f)=0$ for a purely smooth macromodel after removing the macroscopic amplitude, phase, and delay.

\subsection{Inner product} \label{sec:model_innerproduct}

To construct a basis optimized for GW observations, we compare residuals using an inner product in amplification-factor space,
\begin{equation}\label{eq:inner_product}
(F_1|F_2)_A 
=
4\,\Re 
\int_{f_{\min}}^{f_{\max}} df\,
W(f)\,
F_1(f)\,F_2^*(f) ,
\end{equation}
where $W(f)$ is a frequency-dependent weight. 
If two amplification factors act on the same reference waveform $\tilde h_{\rm ref}(f)$, choosing
\begin{equation}
W(f) =
\frac{|\tilde h_{\rm ref}(f)|^2}{S_n(f)}
\end{equation}
recovers the usual detector-noise-weighted GW inner product for the corresponding lensed waveforms.

In this work we adopt a source-agnostic choice and weight only by the detector noise,
\begin{equation}\label{eq:weight_detector}
    W(f) = \frac{\mathcal{N}}{S_n(f)} ,
\end{equation}
where $\mathcal{N}$ is an arbitrary normalization constant. 
This incorporates detector sensitivity without optimizing the basis for a specific binary population. 
The choice of weight matters: stellar microlensing can generate rapid oscillations at high frequencies, and a flat weight would overemphasize features that may be irrelevant for binaries whose observable signal ends at lower frequencies.

More targeted choices are possible. 
For a source population described by parameters $\boldsymbol{\lambda}$, one could define
\begin{equation}\label{eq:weight_population}
 W_{\rm pop}(f)
 \propto
 \int d\boldsymbol{\lambda}\,
 p(\boldsymbol{\lambda})\,
 \frac{|\tilde h(f;\boldsymbol{\lambda})|^2}{S_n(f)} ,
\end{equation}
where $p(\boldsymbol{\lambda})$ encodes the target population. 
This distribution could be inferred empirically from detected compact-binary events~\cite{gwtc4catalog,gwtc5pop}, or obtained from simulated lensed populations that include magnification selection, lensing optical depth, detector sensitivity, and the redshift evolution of the merger rate~\cite{Dai:2016igl,Ng:2017yiu,Oguri:2018muv,Wierda:2021upe,Xu:2021bfn,Li:2026dai}. 
It could also be restricted to a specific detector-frame mass range, producing a basis optimized for the frequencies where those binaries accumulate most of their signal-to-noise. 
For example, a basis tailored to high detector-frame masses would emphasize lower frequencies, while one optimized for lower-mass binaries would retain more high-frequency structure. 
Similarly, a lensed-population prior would weight the basis toward the detector-frame masses and redshifts expected after magnification selection.

Such population-adapted weights may reduce the number of basis elements required for a given class of sources, but they also make the model less generic. 
We therefore use the detector-only weight in Eq.~\eqref{eq:weight_detector} as a conservative baseline and leave source-adapted bases to future work.

We work with discretized normalized weights with frequency values $\{f_i\}$
\begin{equation}\label{eq:weight_normalization}
    w_i(f)
    =
    \frac{W(f_i)}
    {\int_{f_{\min}}^{f_{\max}} df\, W(f)} \,.
\end{equation}
In the \modelname{} model we use the Advanced LIGO zero-detuned high-power noise curve~\cite{LIGOScientific:2014pky,LIGO-T0900288}
to define $S_n(f)$ in Eq.~\eqref{eq:weight_detector}, over the frequency range $f_{\min}=20\,{\rm Hz}$ to $f_{\max}=1024\,{\rm Hz}$. 
This choice defines the ``aLIGO'' label in \modelname. 
The SVD described in Sec.~\ref{sec:model_svd} constructs a basis orthonormal with respect to the discrete version of Eq.~\eqref{eq:inner_product}.

\subsection{Microlensing basis and coefficients} \label{sec:model_svd}

Our goal is to construct a low-dimensional basis for the microlensing residual defined in Eq.~\eqref{eq:delta_F}. 
We write
\begin{equation}\label{eq:microlensing_model}
\delta F(f) \simeq \sum_{k=1}^{K} \alpha_k \psi_k(f) ,
\end{equation}
where $\psi_k(f)$ are basis functions and $\alpha_k$ are phenomenological coefficients. 
Equivalently, the full amplification factor can be written as
\begin{equation}
F(f) \simeq 
F_0(f)
\left[
1 + \sum_{k=1}^{K} \alpha_k \psi_k(f)
\right] ,
\end{equation}
with $F_0(f)$ defined in Eq.~\eqref{eq:F_macro}. 
The basis is obtained from a singular value decomposition (SVD) of a data matrix built from an ensemble of microlensing realizations. 
By construction, it is orthonormal with respect to the discrete version of the inner product in Eq.~\eqref{eq:inner_product}, and ordered by decreasing singular value.

We discretize the residuals on a frequency grid $\{f_j\}_{j=1}^{M}$ and define $w_j\equiv w(f_j)$, where $w(f)$ is the normalized weight in Eq.~\eqref{eq:weight_normalization}.
The simulated amplification factors are computed on a linearly spaced grid with $\Delta f = 1\,{\rm Hz}$; restricting it to the frequency band over which the detector weight is defined, $f\in[20,1024]\,{\rm Hz}$, yields $M=1004$ linearly spaced frequency bins, so the feature vectors defined below have dimension $2M=2008$.
For each microlensing realization, we first compute the residual $\delta F(f)$ by dividing out the smooth macroscopic contribution.
The macroscopic magnification $\mu_{\rm macro}$ is fixed by the external potential, while the constant phase and delay are obtained by fitting the phase of the amplification factor,
\begin{equation}
(\varphi_0,t_0)
=
\underset{\hat\varphi_0,\hat t_0}{\rm argmin}
\sum_{j=1}^{M} w_j
\left[
\arg F(f_j)
-
\left(
\hat\varphi_0 + 2\pi f_j \hat t_0
\right)
\right]^2 .
\label{eq:phase_delay_fit}
\end{equation}
This removes the phase and time shifts that are degenerate with the intrinsic GW parameters.

We represent each residual by concatenating its real and imaginary parts,
\begin{equation}
\mathbf{x} =
\bigl(
\Re\,\delta F_1, \ldots, \Re\,\delta F_M,
\Im\,\delta F_1, \ldots, \Im\,\delta F_M
\bigr) ,
\label{eq:feature_vector}
\end{equation}
where $\delta F_j\equiv \delta F(f_j)$. 
The weighted feature vector is
\begin{align}
\mathbf{x}_w =
\bigl( &
w_1^{1/2}\Re\,\delta F_1, \ldots, w_M^{1/2}\Re\,\delta F_M, \nonumber \\ &
w_1^{1/2}\Im\,\delta F_1, \ldots, w_M^{1/2}\Im\,\delta F_M
\bigr) .
\label{eq:weighted_feature_vector}
\end{align}
With this definition, the Euclidean inner product between weighted feature vectors approximates the weighted inner product of the corresponding complex residuals.

\begin{figure*}
    \includegraphics[width=0.98\textwidth]{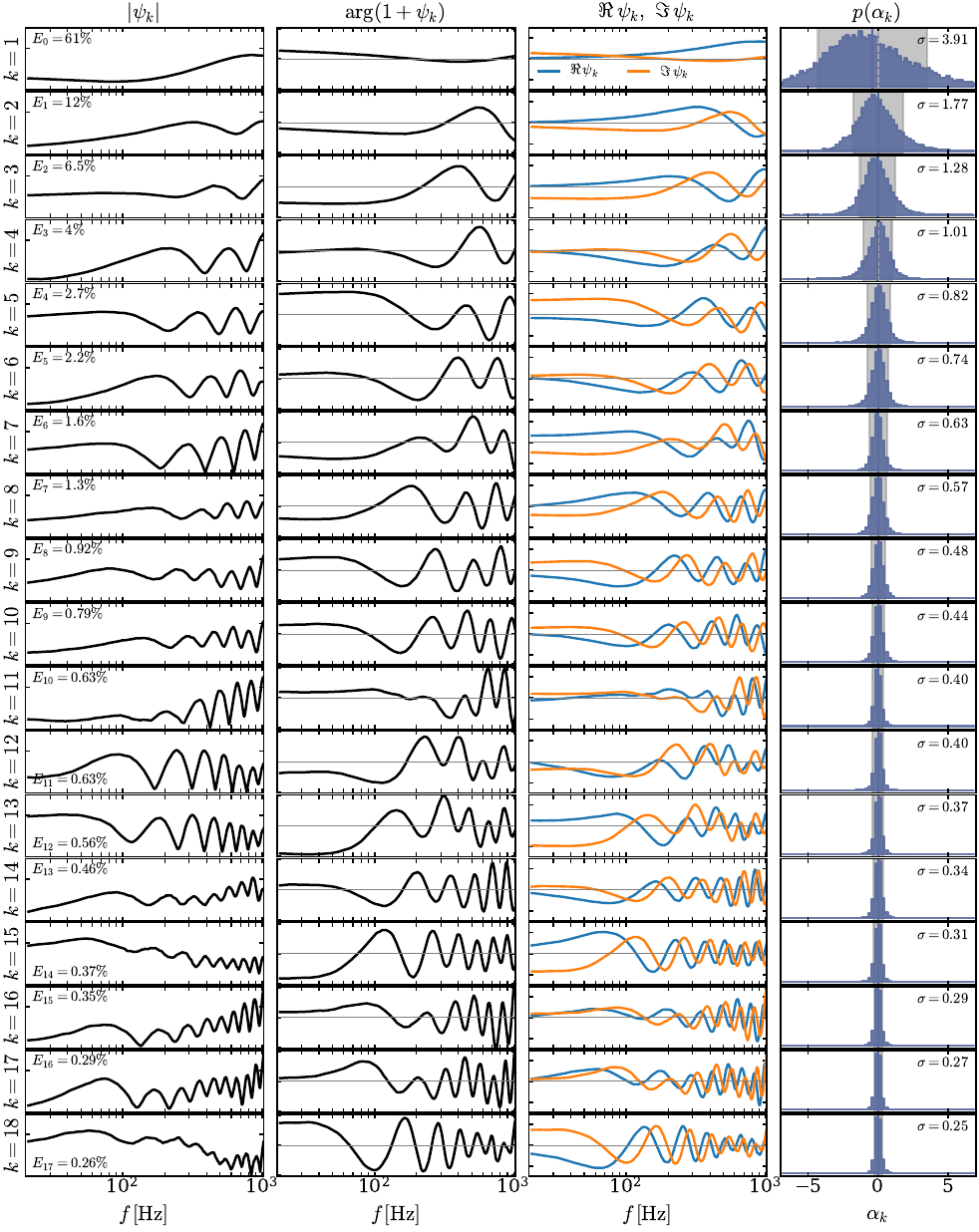}
    \caption{First components of the microlensing basis of the \modelname{} model. 
    The first three columns show the absolute value, argument, and real/imaginary parts of each component, ranked by individual mode power, Eq.~\eqref{eq:individual_power}. 
    No vertical scale is shown because the basis functions are normalized. 
    The last column shows the probability distribution of the coefficients in the microlensing realizations, quoting their standard deviation. 
    The basis was built with a weight proportional to the inverse advanced LIGO noise power spectral density, $W(f)\propto 1/S_n(f)$.}
    \label{fig:microlensing_basis}
\end{figure*}

Given $N$ microlensing realizations, we construct the data matrix
\begin{equation}
X =
\begin{pmatrix}
\mathbf{x}_{w,1} \\
\mathbf{x}_{w,2} \\
\vdots \\
\mathbf{x}_{w,N}
\end{pmatrix}
\in \mathbb{R}^{N \times 2M} ,
\label{eq:data_matrix}
\end{equation}
whose rows contain the weighted feature vectors of the different realizations. 
We compute its singular value decomposition,
\begin{equation}
X = U \Sigma V^{\mathsf T} ,
\label{eq:svd}
\end{equation}
where $\Sigma=\mathrm{diag}(S_1,S_2,\ldots)$ contains the singular values. 
The rows of $V^{\mathsf T}$ define the basis modes in the weighted feature space. 
Equivalently, after undoing the factors of $w_j^{1/2}$, they define complex basis functions $\psi_k(f)$ that are orthonormal under the discrete weighted inner product.

The coefficient of realization $I$ along mode $k$ is
\begin{equation}
\alpha_{k,I}
=
\mathbf{x}_{w,I}\cdot \mathbf{v}_k ,
\label{eq:alpha_projection}
\end{equation}
where $\mathbf{v}_k$ is the $k$-th row of $V^{\mathsf T}$. 
These coefficients encode the diffraction signature of each microlensing realization in the reduced basis. 
Sampling over $\{\alpha_k\}$ therefore provides a phenomenological way to describe microlensing distortions in GW parameter estimation.

The singular values quantify how much of the weighted variance of the training set is captured by each mode. 
We define the individual mode power
\begin{equation}\label{eq:individual_power}
E_k \equiv \frac{S_k^2}{\sum_j S_j^2} ,
\end{equation}
and the cumulative power retained by the first $K$ modes,
\begin{equation}\label{eq:cumulative_power}
E_{\leq K}
\equiv
\frac{\sum_{k=1}^{K} S_k^2}{\sum_j S_j^2}
=
\sum_{k=1}^{K}E_k .
\end{equation}
These quantities measure the variance captured by the basis in the weighted feature space. 
They should be distinguished from the coefficients $\alpha_k$, which describe how strongly a particular microlensing realization excites each mode.

Figure~\ref{fig:microlensing_basis} shows the first components of the basis constructed from 10,000 microlensing realizations. 
The coefficient distributions in the last column show that the first modes have the largest variance, as expected from the ordering by singular value. 
Higher-order modes capture progressively smaller-scale structure in frequency space and are needed to describe more detailed features of individual microlensing realizations.

\section{Microlensing model properties} \label{sec:properties}
The \modelname{} model provides a compact description of microlensing diffraction through the coefficients $\{\alpha_k\}$ in Eq.~\eqref{eq:microlensing_model}. 
These coefficients can be used in two complementary ways:
\begin{itemize}
    \item \textit{Phenomenological inference.} 
    For a chosen truncation order $K$, the coefficients $\{\alpha_k\}_{k=1}^{K}$ can be sampled together with the binary parameters. 
    This allows GW data to determine whether a frequency-dependent microlensing distortion is preferred, without committing to a specific microlens realization or to a detailed astrophysical prior. 
    We demonstrate this wide-prior use of the model on a simulated signal in Sec.~\ref{sec:performance_PE}.

    \item \textit{Astrophysical interpretation.} 
    The ensemble of microlensing realizations used to construct the basis defines a distribution
    \begin{equation}
        p_{\rm realiz}
        \left(
        \{\alpha_k\},\gamma,\kappa_\star,z_L,z_S,\ldots
        \right) .
    \end{equation}
    This distribution encodes the stochastic diffraction signatures expected from stellar fields, including correlations with the macrolens environment. 
    It can be used to define realization-based priors on $\{\alpha_k\}$, as discussed in Sec.~\ref{sec:properties_priors}, or to compare recovered coefficients with the simulated population after parameter estimation. 
    In Sec.~\ref{sec:performance_PE}, we also demonstrate how such realization priors can be applied to posterior samples obtained with wider coefficient priors.
\end{itemize}

In the rest of this section, we introduce simple summary quantities for the coefficient vector (Sec.~\ref{sec:properties_summary_vars}), describe how they correlate with macroscopic lens parameters (Sec.~\ref{sec:properties_macropars}), and define realization-based priors for use in inference (Sec.~\ref{sec:properties_priors}).

\subsection{Microlensing weight and retained power} \label{sec:properties_summary_vars}

Because the basis functions are orthonormal with respect to the weighted inner product, the coefficient vector provides a natural measure of the total microlensing distortion captured by the truncated expansion. 
We define the \emph{microlensing weight} at truncation order $K$ as
\begin{equation}
\upsilon_K \equiv
\left(\sum_{k=1}^{K} \alpha_k^2 \right)^{1/2} .
\label{eq:microlensing_weight}
\end{equation}
This quantity is the weighted norm of the residual retained by the first $K$ basis modes. 
It summarizes the overall amplitude of the microlensing distortion, while being insensitive to the signs of individual coefficients, which depend on the detailed interference pattern of each realization.

To quantify the effect of truncating the expansion, we define the retained weight fraction
\begin{equation}
r_K \equiv \frac{\upsilon_K}{\upsilon_{K_{\max}}} ,
\label{eq:captured_power}
\end{equation}
where $K_{\max}$ is the largest basis size used as reference. 
This satisfies $0\le r_K\le1$. 
Values close to unity indicate that the first $K$ modes capture most of the microlensing distortion present in the reference basis, while smaller values indicate that higher-order modes contribute significantly.

We use $\upsilon_K$ below as a compact statistic for the strength of microlensing diffraction, and $r_K$ in Sec.~\ref{sec:performance} to assess the accuracy of truncated reconstructions.

\subsection{Dependence on macro-parameters} \label{sec:properties_macropars}

The coefficient distribution contains information about the macroscopic lens environment. 
In particular, the ensemble of realizations defines conditional distributions such as
\begin{equation}
    p_{\rm realiz}
    \left(
    \{\alpha_k\}\mid \gamma,\kappa_\star,z_L,z_S,\ldots
    \right) ,
\end{equation}
where $\gamma$ is the external shear and $\kappa_\star$ is the stellar convergence. 
Individual coefficients are not expected to provide clean estimators of these quantities, since their signs and amplitudes depend on the detailed interference pattern of each stellar field. 
However, summary statistics such as $\upsilon_K$ are more robust: the macrolens parameters mainly control the probability of producing strong microlensing distortions, rather than the exact coefficient vector of a single realization.

We therefore examine the correlation between the microlensing weight $\upsilon_K$, defined in Eq.~\eqref{eq:microlensing_weight}, and the macrolens parameters. 
This provides a simple measure of how the overall strength of diffraction depends on the lens environment. 
More generally, one could construct optimized estimators for $\gamma$, $\kappa_\star$, $z_L$, or $z_S$ from the full coefficient vector $\{\alpha_k\}$; we leave this to future work.

\begin{figure}
    \centering
    \includegraphics[width=0.99\linewidth]{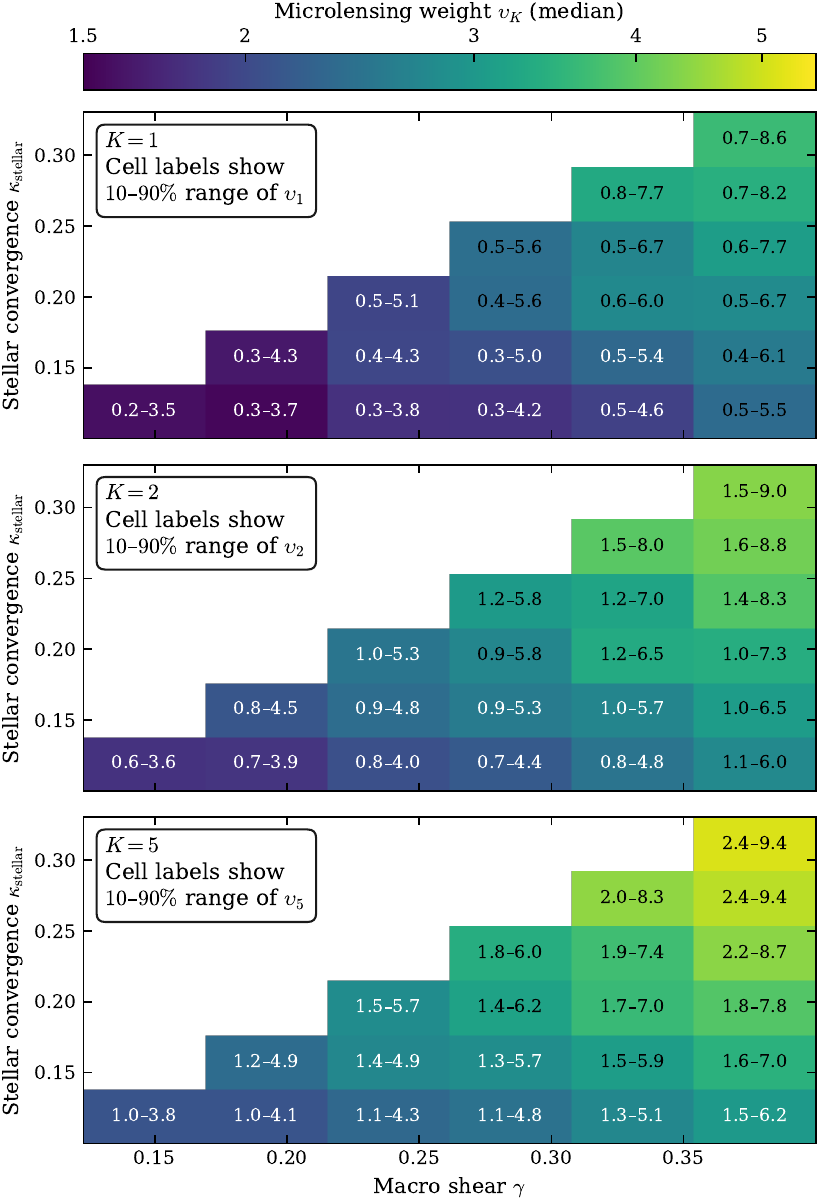}
    \caption{Microlensing weight $\upsilon_K$ in the $(\gamma,\kappa_{\rm stellar})$ plane for the realizations used in \modelname, at truncation orders $K=1,2,$ and $5$  (top to bottom).
    Colors show the median value of $\upsilon_K$ computed from the ensemble of microlensing realizations within each bin of macro shear $\gamma$ and stellar convergence $\kappa_{\rm stellar}$, related to the total compact-object convergence by $\kappa_\star=1.2\,\kappa_{\rm stellar}$.
    Numbers inside the cells indicate the $10$--$90\%$ range of $\upsilon_K$ within each bin. 
    The color scale is shared across panels to allow direct comparison of how the recovered microlensing weight increases as additional basis modes are included.}
    \label{fig:weights_summary_gamma_kappastar}
\end{figure}

Figure~\ref{fig:weights_summary_gamma_kappastar} shows that the microlensing weight increases with both shear and stellar convergence. 
This behavior is expected: $\kappa_\star$ controls the surface density of compact lenses, while $\gamma$ changes the local tidal field and increases the region over which microlenses can produce large perturbations. 
The broad $10$--$90\%$ ranges reflect the stochasticity of the stellar field. 
The macro-parameters determine a distribution of possible diffraction strengths, not a unique value of $\upsilon_K$ for a single realization.

\begin{figure}
    \centering
    \includegraphics[width=0.99\linewidth]{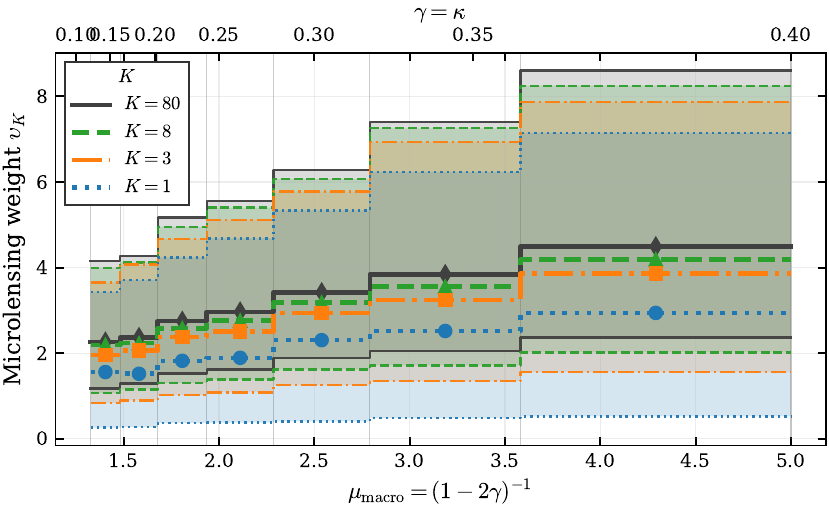}
    \caption{
    Microlensing weight as a function of macromagnification for \modelname. 
    Curves show the median $\upsilon_K$ for truncation orders $K=1,3,8,80$, while shaded bands indicate the $10$--$90\%$ range in bins of $\gamma$. 
    The lower axis uses the SIS relation in Eq.~\eqref{eq:mu_macro}; the upper axis shows the equivalent value of $\gamma$.}
    \label{fig:upsilon_vs_mu}
\end{figure}

Figure~\ref{fig:upsilon_vs_mu} shows the same trend as a function of macromagnification.
The microlensing weight increases systematically with $\mu_{\rm macro}$ for all truncation orders.
Increasing $K$ raises the recovered weight, as expected from Eq.~\eqref{eq:microlensing_weight}, but the dependence on $\mu_{\rm macro}$ remains qualitatively similar.
This indicates that the correlation is not driven by a single mode, but by the overall strength of the microlensing residual.

\begin{figure}
    \centering
    \includegraphics[width=0.99\linewidth]{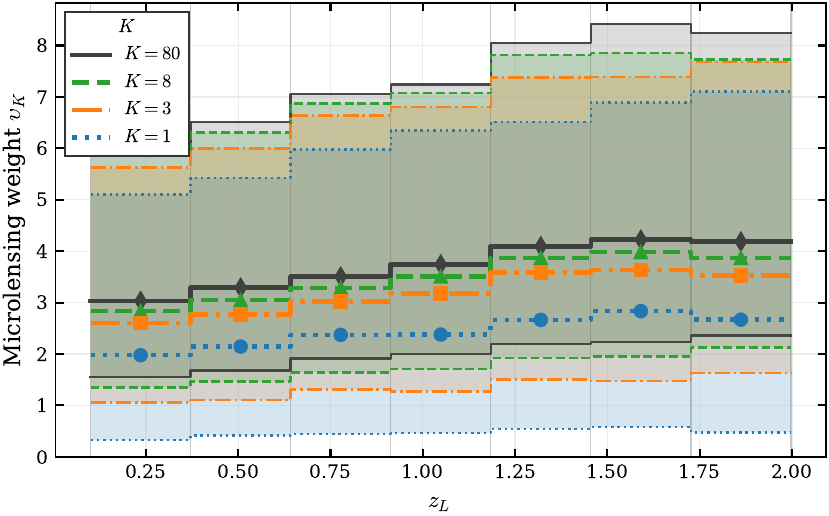}
    \caption{Microlensing weight as a function of lens redshift for \modelname{}.
    Curves show the median $\upsilon_K$ for truncation orders $K=1,3,8,80$, while shaded bands indicate the $10$--$90\%$ range in bins of $z_L$.
    The mild increase with $z_L$ reflects the growth of the redshifted microlens masses, $\propto(1+z_L)$, which shifts diffraction features toward lower observed frequencies within the sensitive band.
    The trend is subdominant compared to the dependence on $\gamma$ and $\kappa_\star$ (Figs.~\ref{fig:weights_summary_gamma_kappastar} and \ref{fig:upsilon_vs_mu}) and to the realization-to-realization scatter.}
    \label{fig:upsilon_vs_zL}
\end{figure}

The dependence of the microlensing weight on the lens and source redshifts is comparatively weak.
At fixed $\kappa$, $\gamma$, and $\kappa_\star$, the source redshift does not alter the statistics of the amplification factor: the microlens surface density is fixed by the stellar convergence, and $z_S$ enters only the conversion between physical and dimensionless quantities.
The redshifts affect the observed signal mainly through the redshifted microlens masses, $M_j(1+z_L)$, which shift the diffraction pattern toward lower observed frequencies as $z_L$ increases, moving spectral features across the band selected by the weight $w(f)$.
Figure~\ref{fig:upsilon_vs_zL} shows the resulting trend in the simulated ensemble: the median $\upsilon_K$ increases mildly with $z_L$, by ${\sim}30\%$ between the lowest and highest redshift bins for $K=8$.
The rank correlation between $\upsilon_8$ and $z_L$ is $\rho\simeq0.20$, roughly half of that with the shear ($\rho\simeq0.35$) or the stellar convergence ($\rho\simeq0.38$), and remains at a similar level within narrow bins of $\gamma$ or $\kappa_\star$, confirming that it is a genuine, but subdominant, dependence rather than one mediated by the sampled macro-parameters.
We note that in an astrophysical lens population the redshifts would additionally influence the microlensing signal through $\kappa_\star$ itself, via the critical surface density $\Sigma_{\rm cr}(z_L,z_S)$; in our ensemble this channel is absent by construction, since $\kappa_\star$ is an independent input parameter.

These correlations suggest how measured coefficients could be interpreted astrophysically.
A posterior lower bound on $\upsilon_K$ from a GW event would disfavor regions of parameter space where such large weights are rare, leading to probabilistic constraints on $\gamma$, $\kappa_\star$, and derived quantities such as $\mu_{\rm macro}$. 
Equivalently, one can compare posterior samples for $\{\alpha_k\}$ directly with the simulated distribution 
$p_{\rm realiz}(\{\alpha_k\}\mid\gamma,\kappa_\star,z_L,z_S)$, or with compressed summaries such as 
$p_{\rm realiz}(\upsilon_K\mid\gamma,\kappa_\star)$. 
This provides a calibrated, stochastic interpretation of the recovered microlensing distortion, rather than a deterministic reconstruction of the macrolens parameters.

\subsection{Realization-based priors} \label{sec:properties_priors}

The same ensemble of simulated amplification factors used to construct the basis also defines an empirical distribution for the coefficients $\{\alpha_k\}$. 
Let
\begin{equation}
    \boldsymbol{\lambda}
    =
    (\gamma,\kappa_\star,z_L,z_S,\ldots)
\end{equation}
denote the macroscopic and population-level variables that specify a microlensing realization. 
For a basis truncated at order $K$, we define
\begin{equation}
    \boldsymbol{\alpha}
    =
    (\alpha_1,\ldots,\alpha_K) .
\end{equation}
The realization ensemble induces the marginal coefficient distribution
\begin{equation}
    p_{\rm realiz}^{(K)}(\boldsymbol{\alpha})
    =
    \int d\boldsymbol{\lambda}\,
    p(\boldsymbol{\alpha}\mid\boldsymbol{\lambda})\,
    p_{\rm realiz}(\boldsymbol{\lambda}) ,
    \label{eq:realization_prior_def}
\end{equation}
where $p_{\rm realiz}(\boldsymbol{\lambda})$ is the sampling distribution used to generate the simulations. 
This is not, in general, an astrophysical event-rate prior. 
Rather, it is a prior induced by the agnostically sampled realization ensemble.

In the simplest implementation, we approximate this distribution by a multivariate Gaussian,
\begin{equation}
    p_{\rm realiz}^{(K)}(\boldsymbol{\alpha})
    =
    \mathcal{N}
    \left(
    \boldsymbol{\alpha};
    \boldsymbol{\mu}_{\alpha},
    \boldsymbol{\Sigma}_{\alpha}
    \right),
    \label{eq:realization_prior_gaussian}
\end{equation}
with mean and covariance estimated from the simulated coefficient samples,
\begin{align}
    \boldsymbol{\mu}_{\alpha}
    &=
    \left\langle \boldsymbol{\alpha}_I \right\rangle_I ,
    \\
    \boldsymbol{\Sigma}_{\alpha}
    &=
    \left\langle
    \left(\boldsymbol{\alpha}_I-\boldsymbol{\mu}_{\alpha}\right)
    \left(\boldsymbol{\alpha}_I-\boldsymbol{\mu}_{\alpha}\right)^{\mathsf T}
    \right\rangle_I ,
    \label{eq:realization_prior_cov}
\end{align}
where $I$ labels the microlensing realization. 
Because the microlensing basis diagonalizes the covariance of the weighted training set, this Gaussian approximation is close to diagonal when estimated from the same ensemble used to construct the basis. 
Non-Gaussian features, correlations induced by conditioning, and tails associated with rare high-diffraction realizations are not captured by this approximation.

The realization prior can also be conditioned on properties of the simulated lensing configuration. 
For a subset $\mathcal{S}$ of realizations, for example a cut on shear, stellar convergence, macromagnification, or image type, one can define
\begin{equation}
    p_{\rm realiz}^{(K)}(\boldsymbol{\alpha}\mid \mathcal{S})
    =
    p_{\rm realiz}(\alpha_1,\ldots,\alpha_K\mid \mathcal{S}) .
    \label{eq:sliced_prior}
\end{equation}
This provides a practical way to construct ``sliced'' priors associated with different regions of the simulated microlensing population. 
In this work we use only the unconditional realization prior as a first demonstration. 
More flexible density estimates, such as kernel-density estimates, Gaussian mixtures, or normalizing flows, can be substituted for Eq.~\eqref{eq:realization_prior_gaussian} without changing the inference strategy.

A realization prior can be applied directly in parameter estimation, or by reweighting posterior samples obtained with wider coefficient priors. 
If the sampling prior used in parameter estimation is $p_{\rm PE}(\boldsymbol{\alpha})$, then the realization-prior posterior is obtained by assigning each posterior sample a weight
\begin{equation}
    w_{\rm post}
    \propto
    \frac{
        p_{\rm realiz}^{(K)}(\boldsymbol{\alpha})
    }{
        p_{\rm PE}(\boldsymbol{\alpha})
    } .
    \label{eq:posterior_reweighting_realiz}
\end{equation}
This strategy is demonstrated in Sec.~\ref{sec:performance_PE}, where we first sample the microlensing coefficients with broad priors and then apply the realization-based prior by reweighting.

Finally, the realization distribution can be replaced by an astrophysical population model. 
If the simulations are drawn from a known distribution $p_{\rm realiz}(\boldsymbol{\lambda})$, and a target population model $p_{\rm astro}(\boldsymbol{\lambda})$ is specified, the coefficient distribution can be reweighted at the realization level using
\begin{equation}
    w_I
    \propto
    \frac{
        p_{\rm astro}(\boldsymbol{\lambda}_I)
    }{
        p_{\rm realiz}(\boldsymbol{\lambda}_I)
    } .
    \label{eq:astro_reweighting}
\end{equation}
The corresponding astrophysical coefficient prior is then approximated by
\begin{equation}
    p_{\rm astro}^{(K)}(\boldsymbol{\alpha})
    \simeq
    \sum_I
    w_I\,
    \delta^{(K)}
    \left(
    \boldsymbol{\alpha}-\boldsymbol{\alpha}_I
    \right) .
    \label{eq:astro_prior_empirical}
\end{equation}
We do not perform this astrophysical reweighting here, but Eq.~\eqref{eq:astro_reweighting} shows how the same coefficient framework can be adapted to different assumptions about the lens population.

\section{Microlensing model performance} \label{sec:performance}

We now assess how effectively  microlensing diffraction is captured by the \modelname{} model. 
Since the basis is complete only in the limit of sufficiently many modes, the practical question is how many coefficients are needed for the applications of interest. 
We first quantify the effect of truncating the basis on the recovered microlensing weight (Sec.~\ref{sec:performance_truncation}). 
We then demonstrate the use of the model in Bayesian inference on a simulated microlensed signal (Sec.~\ref{sec:performance_PE}). 
Finally, we apply the same basis to an isolated point lens (Sec.~\ref{sec:performance_PL}), which lies outside the stochastic stellar-field population used to construct the model and therefore provides an out-of-distribution test.

\subsection{Truncation of the microlensing basis}\label{sec:performance_truncation}

We first quantify how the performance of the \modelname{} basis depends on the truncation order $K$. 
For each realization, we compute the microlensing weight $\upsilon_K$, defined in Eq.~\eqref{eq:microlensing_weight}, and compare it with the value obtained using the largest basis considered here, $K_{\max}=80$. 
We use the retained weight fraction $r_K$, Eq.~\eqref{eq:captured_power}, to measure how much of the weighted microlensing norm is captured by the first $K$ modes.

\begin{figure}[t]
    \centering
    \includegraphics[width=0.99\linewidth]{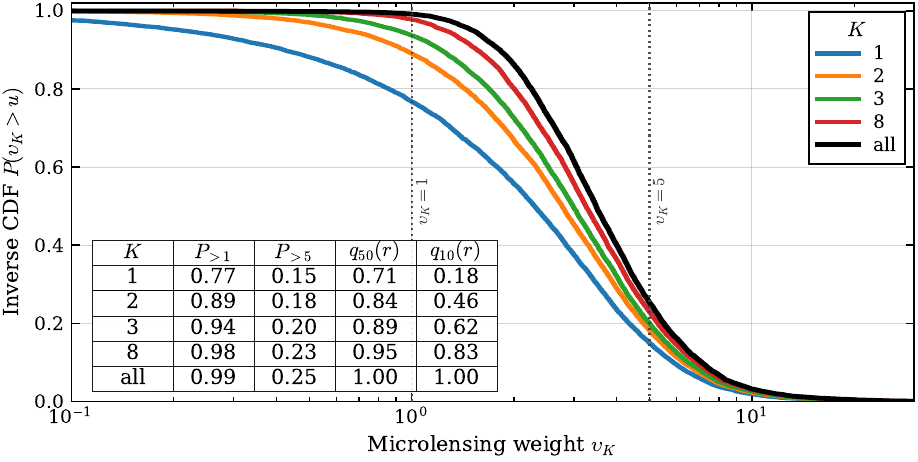}
    \caption{
    Inverse cumulative distribution of the microlensing weight $\upsilon_K$ for different truncation orders in the \modelname{} realizations. 
    The full reference basis corresponds to $K=80$. 
    Vertical dotted lines mark the reference amplitudes $\upsilon_K=1$ and $\upsilon_K=5$. 
    The inset reports the fraction of realizations above these thresholds, together with the median and 10th percentile of the retained weight fraction $r_K=\upsilon_K/\upsilon_{80}$.
    }
    \label{fig:truncation_weight}
\end{figure}

Figure~\ref{fig:truncation_weight} shows the inverse cumulative distribution of $\upsilon_K$ for several truncation orders. 
The first few modes already retain most of the microlensing weight for the majority of realizations. 
For example, $K=3$ gives a median retained fraction $q_{50}(r_K)\simeq0.89$, while $K=8$ increases this to $q_{50}(r_K)\simeq0.95$. 
The lower tail is more sensitive to truncation: the 10th percentile increases from $q_{10}(r_K)\simeq0.62$ for $K=3$ to $q_{10}(r_K)\simeq0.83$ for $K=8$. 
Thus a modest number of coefficients captures the dominant microlensing structure for most realizations, while higher-order modes remain relevant for the most complex diffraction patterns.

The abundance of large microlensing weights is also reasonably stable once several modes are included. 
The fraction of realizations with $\upsilon_K>1$ is already close to the full-basis value for $K=3$--$8$. 
The more stringent threshold $\upsilon_K>5$ is more sensitive to truncation, as expected because large distortions often receive contributions from several subdominant modes.

In the following subsections we use $K=8$ as a representative truncation. 
This choice is not unique, nor is it guaranteed to be optimal for all inference problems. 
It provides a compact model that retains most of the microlensing norm for the bulk of the realization ensemble, while keeping the number of additional parameters low enough for parameter-estimation studies. 
The dependence of the inference results on $K$ should be assessed in future applications, especially for high-SNR events or signals with unusually complex diffraction structure.

\subsection{Bayesian parameter estimation} \label{sec:performance_PE}

We now demonstrate how the \modelname{} model can be used in Bayesian parameter estimation.
The purpose of this example is twofold.
First, we show that the microlensing coefficients can be sampled with broad priors as a flexible phenomenological model for diffraction.
Second, we show how the realization-based coefficient prior of Sec.~\ref{sec:properties_priors} can be applied a posteriori by reweighting the broad-prior posterior samples.

\begin{figure*}[t]
    \centering
    \includegraphics[width=0.86\textwidth]{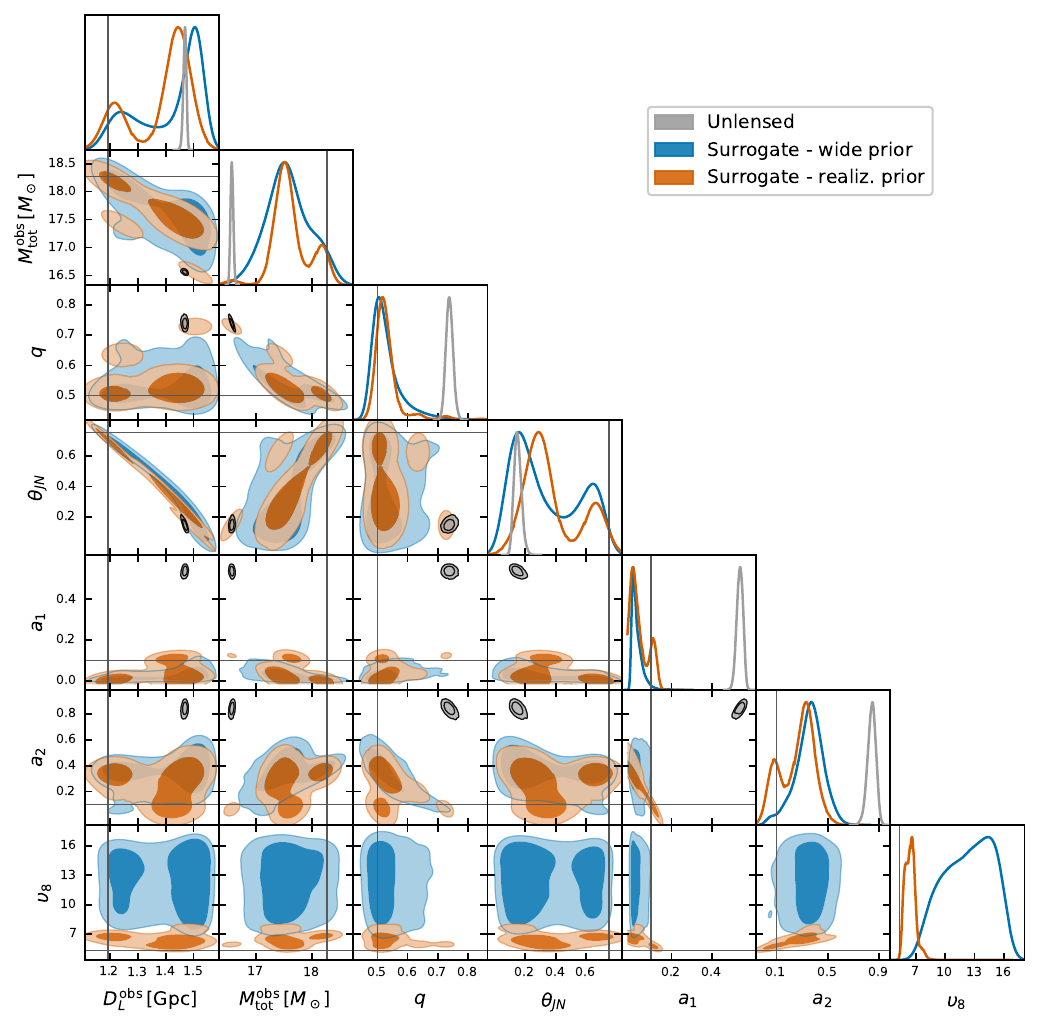}
    \caption{
    Source-parameter recovery of an injected microlensed signal.
    We compare the unlensed recovery, the recovery with wide priors on the microlensing coefficients using \modelname{}, and the same posterior reweighted by the realization-based coefficient prior.
    The wide-prior microlensing posterior captures the microlensing distortion, but allows large microlensing weights $\upsilon_8$ and broader degeneracies with the source parameters.
    Reweighting by the realization prior restricts the posterior to coefficient combinations typical of the simulated stellar fields, reducing the recovered microlensing weight toward the projected injected value.
    Thin gray lines indicate injected source-parameter values and the microlensing weight obtained by projecting the true injected amplification factor onto the first eight basis modes.
    }
    \label{fig:source_weight_triangle}
\end{figure*}

\begin{figure*}[t]
    \centering
    \includegraphics[width=0.99\textwidth]{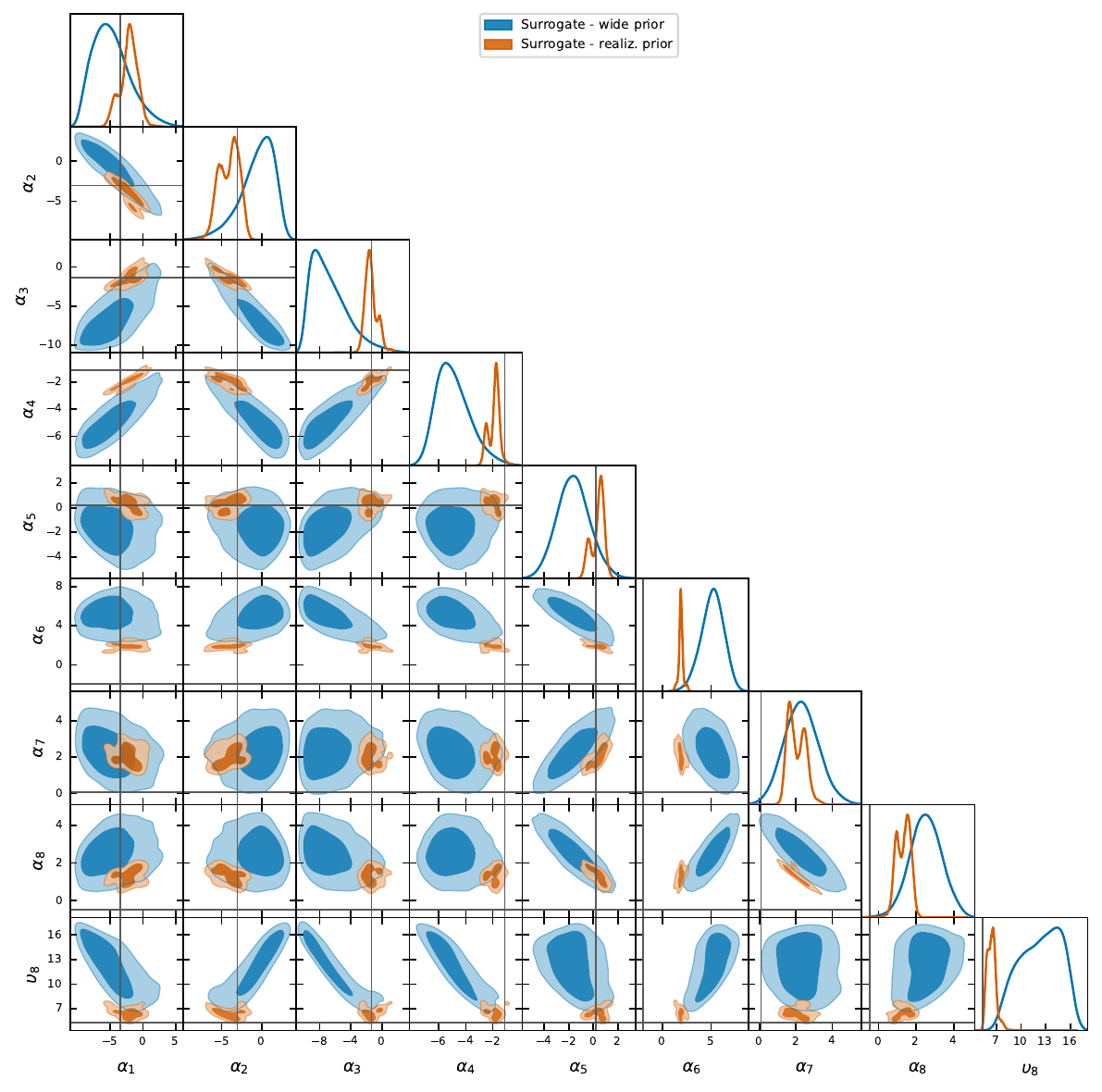}
    \caption{
    Recovery of the microlensing coefficients for the injected microlensed signal.
    The coefficients shown correspond to the first eight basis modes of \modelname{}.
    The wide-prior posterior explores broad and correlated regions of coefficient space.
    Reweighting by the realization prior selects the subset of this posterior that is typical of the simulated stellar-field realizations.
    Thin gray lines indicate the values obtained by projecting the full injected amplification factor onto the first eight basis modes.
    }
    \label{fig:svd_coefficients_triangle}
\end{figure*}

\begin{table}[t]
    \centering
    \begingroup
    \renewcommand{\arraystretch}{1.15}
    \setlength{\tabcolsep}{4pt}
    \begin{tabular}{lcccc}
    \toprule
    Parameter & Injection & Unlensed & Wide prior & Realiz. prior \\
    \midrule
    $D_L^{\rm obs}\,[{\rm Gpc}]$ 
        & $1.19$ 
        & $1.47^{+0.01}_{-0.01}$ 
        & $1.48^{+0.05}_{-0.22}$ 
        & $1.41^{+0.05}_{-0.20}$ \\
    $M_{\rm tot}^{\rm obs}\,[M_\odot]$ 
        & $18.3$ 
        & $16.6^{+0.0}_{-0.0}$ 
        & $17.6^{+0.5}_{-0.4}$ 
        & $17.6^{+0.5}_{-0.3}$ \\
    $q$ 
        & $0.50$ 
        & $0.74^{+0.01}_{-0.01}$ 
        & $0.52^{+0.06}_{-0.03}$ 
        & $0.52^{+0.04}_{-0.02}$ \\
    $\theta_{JN}$ 
        & $0.75$ 
        & $0.15^{+0.03}_{-0.02}$ 
        & $0.27^{+0.36}_{-0.16}$ 
        & $0.36^{+0.30}_{-0.10}$ \\
    $a_1$ 
        & $0.10$ 
        & $0.54^{+0.02}_{-0.02}$ 
        & $0.02^{+0.03}_{-0.01}$ 
        & $0.03^{+0.08}_{-0.03}$ \\
    $a_2$ 
        & $0.10$ 
        & $0.84^{+0.03}_{-0.03}$ 
        & $0.36^{+0.09}_{-0.11}$ 
        & $0.32^{+0.03}_{-0.25}$ \\
    \midrule
    $\upsilon_{8}$ 
        & $5.32$ 
        & -- 
        & $12.6^{+2.5}_{-3.1}$ 
        & $6.5^{+0.4}_{-0.7}$ \\
    \bottomrule
    \end{tabular}
    \endgroup
    \caption{
    Summary of one-dimensional posterior constraints for the injected microlensed signal.
    The first column gives the injected values, while the remaining columns show the median and $68\%$ credible intervals for the unlensed recovery, the \modelname{} recovery with wide coefficient priors, and the same posterior reweighted by the realization-based coefficient prior.
    The microlensing weight $\upsilon_8$ is computed from the first eight microlensing coefficients.
    }
    \label{tab:table_PE}
\end{table}

The injected signal is generated using the full numerical amplification factor $F(f)$ from one microlensing realization, rather than from the truncated reduced-order model itself.
This makes the test nontrivial: the recovery model has access only to the first $K=8$ basis components, while the injected waveform contains the full diffraction pattern of the simulated stellar field.
The injected source parameters are quoted in Table~\ref{tab:table_PE} and the lensing configuration is $\kappa=\gamma=0.42$, $\kappa_\star=0.28$, $z_L=1.14$, $z_S=1.83$, $\mu_{\rm macro}=6.33$.
This configuration lies slightly outside the range sampled to construct the basis ($\kappa=\gamma\leq0.4$, corresponding to $\mu_{\rm macro}\leq5$).
The choice is deliberate: it stress-tests the robustness of the model under mild extrapolation, which is representative of realistic applications where the macro-parameters of an observed event are unknown and not guaranteed to lie within the training range.
The smooth dependence of the coefficient statistics on the macro-parameters (Figs.~\ref{fig:weights_summary_gamma_kappastar} and \ref{fig:upsilon_vs_mu}) and their weak dependence on redshift (Fig.~\ref{fig:upsilon_vs_zL}) suggest that the basis remains adequate slightly beyond the training boundary; the successful recovery presented below confirms this expectation, and injections within the training range are expected to perform at least as well.

The reduced-order model does not include the external macromagnification $\mu_{\rm macro}$ as an inferred parameter.
It therefore measures apparent, or observed, source quantities rather than the intrinsic unlensed luminosity distance, redshift, and source-frame masses.
We quote $D_L^{\rm obs}=D_L^{\rm true}/\sqrt{\mu_{\rm macro}}$.
The corresponding observed redshift $z_{\rm obs}$ is inferred from $D_L^{\rm obs}$ assuming a $\Lambda$CDM cosmology~\cite{Planck:2018lbu}, and the observed total mass is
$M_{\rm tot}^{\rm obs}=M_{\rm tot}^{\rm det}/(1+z_{\rm obs})
= M_{\rm tot}^{\rm true}(1+z_{\rm true})/(1+z_{\rm obs})$.

The injection--recovery analyses are performed with \textsc{Bilby} using the \textsc{dynesty} nested sampler.
We use a three-detector network consisting of third-generation GW detectors. Specifically, the network comprises two Cosmic Explorer~\citep{2019BAAS...51g..35R} detectors located at Hanford and Livingston, and one Einstein Telescope~\citep{2012CQGra..29l4013S} detector near Pisa, Italy.
The power spectral densities (PSDs) employed in this work are CosmicExplorerP1600143 and EinsteinTelescopeP1600143 from Ref.~\citep{2017CQGra..34d4001A}.
It is worth noting that, as the first paper in this series, we focus on a sufficiently high-SNR event, with a network SNR of approximately 450, as a deliberately optimistic demonstration to provide a proof of principle for the feasibility of the SVD method.
In future work, we will investigate the detection efficiency, degeneracies with source parameters, and related aspects under the current LIGO-Virgo-KAGRA (LVK) sensitivity.

We employ a frequency-domain likelihood with time and distance marginalization enabled, while phase and calibration marginalization are not included.

The GW signal is generated with the precessing waveform approximant \texttt{IMRPhenomXP}~\cite{Pratten:2020ceb}, using a reference frequency $f_{\rm ref}=50\,{\rm Hz}$ and a minimum frequency $f_{\min}=20\,{\rm Hz}$.
The data are sampled at $4096\,{\rm Hz}$ over a duration of $3.84\,{\rm s}$.
The source parameters quoted in Table~\ref{tab:table_PE} are the observed luminosity distance $D_L^{\rm obs}$ and total mass $M_{\rm tot}^{\rm obs}$ defined above, the mass ratio $q=m_2/m_1\leq1$, the angle $\theta_{JN}$ between the total angular momentum and the line of sight, and the dimensionless spin magnitudes $a_1$ and $a_2$ of the primary and secondary components.
The lensed recovery uses a frequency-domain source model in which the compact-binary waveform is multiplied by the microlensing amplification factor reconstructed from \modelname{}.

Note that the microlensing basis itself was constructed with the aLIGO noise curve as normalized weight, rather than the PSDs used in this injection.
This is slightly sub-optimal for the present PE example, but it is useful as a proof of principle; a detector-specific basis can be constructed by replacing the weight in Eq.~\eqref{eq:weight_detector}.

For comparison with the recovered microlensing parameters, we compute fiducial coefficient values by projecting the injected microlensing residual onto the first eight basis components,
\begin{equation}
    \alpha_i^{\rm fid}
    =
    \left(\delta F_{\rm inj}\mid \psi_i\right),
    \qquad i=1,\ldots,8 .
    \label{eq:fiducial_alpha_PE}
\end{equation}
These values define the gray lines in Fig.~\ref{fig:svd_coefficients_triangle}.
They also determine the projected injected value of $\upsilon_8$ shown in Fig.~\ref{fig:source_weight_triangle} and quoted in Table~\ref{tab:table_PE}.
The fiducial coefficients are not parameters used to generate the signal; they are the projection of the full injected amplification factor onto the truncated recovery basis.

We compare three analyses of the same microlensed signal.
First, we analyze the data with an unlensed waveform model.
Second, we use the \modelname{} model with wide, uniform priors on the first eight microlensing coefficients.
Third, we reweight the wide-prior microlensing posterior by the realization prior introduced in Sec.~\ref{sec:properties_priors}.
For the wide-prior run, each coefficient is sampled with $\alpha_i\in[-10,10]$, $i=1,\ldots,8$.
These deliberately broad coefficient priors provide a conservative stress test of the microlensing recovery, but they also introduce a large prior volume and make posterior reweighting inefficient.

The reweighting is performed a posteriori using samples from the wide-prior microlensing run.
For posterior samples drawn from
$p_{\rm wide}(\boldsymbol{\theta},\boldsymbol{\alpha}\mid d)$,
the realization-prior posterior is
\begin{equation}
    p_{\rm realiz}(\boldsymbol{\theta},\boldsymbol{\alpha}\mid d)
    \propto
    p_{\rm wide}(\boldsymbol{\theta},\boldsymbol{\alpha}\mid d)
    \frac{
    \pi_{\rm realiz}^{(K)}(\boldsymbol{\alpha})
    }{
    \pi_{\rm wide}^{(K)}(\boldsymbol{\alpha})
    },
    \label{eq:PE_reweighting_posterior}
\end{equation}
where $\boldsymbol{\theta}$ denotes the binary parameters and $d$ the data.
Equivalently, each wide-prior posterior sample is assigned an importance weight
\begin{equation}
    w_i \propto
    \frac{
    \pi_{\rm realiz}^{(K)}(\boldsymbol{\alpha}_i)
    }{
    \pi_{\rm wide}^{(K)}(\boldsymbol{\alpha}_i)
    } .
    \label{eq:PE_reweighting_weights}
\end{equation}
Since $\pi_{\rm wide}^{(K)}$ is constant inside the sampled prior volume, the weights are proportional to
$\pi_{\rm realiz}^{(K)}(\boldsymbol{\alpha}_i)$.
As in Sec.~\ref{sec:properties_priors}, this realization prior is derived from the agnostically sampled microlensing ensemble; it is not an astrophysical event-rate prior.
The Gaussian realization prior used here is estimated from the coefficient matrix of the $10^4$ simulated stellar-field realizations used to build the basis.

Figure~\ref{fig:source_weight_triangle} shows the posterior on selected source parameters and on the recovered microlensing weight.
The unlensed analysis cannot reproduce the injected diffraction pattern and compensates through biased source parameters, including the observed luminosity distance, component spins, and mass ratio.
The wide-prior microlensing model captures the signal more flexibly and improves several intrinsic parameters, in particular the mass ratio, spin magnitudes, and observed total mass.
However, the broad coefficient prior also allows unrealistically large microlensing distortions: the posterior peaks at $\upsilon_8$ values significantly above the projected injected value.
After reweighting by the realization prior, the posterior is restricted to coefficient combinations that are typical of the simulated stellar fields.
This reduces the inferred microlensing weight from $\upsilon_8=12.6^{+2.5}_{-3.1}$ to $\upsilon_8=6.5^{+0.4}_{-0.7}$, closer to the projected value $\upsilon_8=5.32$.

Figure~\ref{fig:svd_coefficients_triangle} shows the corresponding posterior for the microlensing  coefficients.
With wide priors, the posterior explores broad and correlated directions in coefficient space.
These correlations reflect degeneracies among basis components and between lensing and source parameters.
The realization prior selects a much smaller region of the wide-prior posterior, concentrating the recovery near coefficient combinations that occur in the simulated microlensing ensemble.
While the posterior of the first basis coefficients ($\alpha_1$--$\alpha_5$) are drawn close to the injection values, the last three coefficients ($\alpha_6$--$\alpha_8$) remain far from their expected fiducial. Although the base elements are orthogonal in the amplification factor $\delta F$, correlations with the source parameters are likely to introduce these biases. These correlations are realization-dependent: they depend on the injected diffraction pattern, the source parameters, and the detector sensitivity. 
Consequently, the optimal truncation order $K$ is not fixed only by the retained microlensing weight in the training set, but also by the balance between improved waveform flexibility and the additional prior volume introduced in parameter estimation.

The same importance weights can be used to estimate the evidence for each analysis, 
$Z_{\rm realiz}=Z_{\rm wide}\left\langle \pi_{\rm realiz}^{(K)}(\boldsymbol{\alpha})/\pi_{\rm wide}^{(K)}(\boldsymbol{\alpha})\right\rangle_{\rm wide}$, 
where the average is over posterior samples from the wide-prior microlensing analysis.
For this injection we find
\begin{align}
    \log_{10} \mathcal{B}_{\rm micro,wide}^{\rm UL}
    &= 40.3 ,
    \\
    \log_{10} \mathcal{B}_{\rm micro,realiz}^{\rm UL}
    &= 36.4 ,
\end{align}
with Bayes factors quoted relative to the unlensed model.
Thus both microlensing analyses are decisively preferred over the unlensed recovery, but in this case the realization prior reduces the evidence relative to the wide-prior microlensing model.
This is not a generic consequence of using a narrower prior: it occurs because the wide-prior posterior places substantial support on coefficient combinations, and in particular large $\upsilon_8$, that are atypical of the simulated stellar-field realizations.

The reweighting diagnostics indicate that the realization-prior posterior is sampled with very low efficiency.
The effective reweighting efficiency is $\epsilon_{\rm rw}\equiv N_{\rm eff}/N_{\rm samp}\simeq4.0\times10^{-4}$.
Only a tiny fraction of the wide-prior posterior samples contribute appreciably after applying the realization prior: the samples carrying $50\%$, $90\%$, and $99\%$ of the reweighted posterior are only $3$, $10$, and $40$, respectively.
This concentration is also reflected in the large spread of importance weights, $\log_{10}[\max_i w_i/\mathrm{median}(w_i)]\simeq26.2$.
The low efficiency is expected because the realization prior occupies only a small fraction of the very broad coefficient-prior volume.
For $K$ coefficients, the relevant volume contraction scales roughly as $(\sigma_i/\Delta\alpha_i)^K$, where $\sigma_i$ is the typical realization-prior width and $\Delta\alpha_i$ is the flat-prior width.
This sparsity also explains the irregular shape of the reweighted contours in Figs.~\ref{fig:source_weight_triangle} and \ref{fig:svd_coefficients_triangle}.

The reweighted posterior is therefore useful as a proof of principle, but a more robust comparison should use dedicated sampling with the realization prior imposed directly.
In practice, the need for dedicated sampling can be anticipated from the prior-volume contraction, or diagnosed a posteriori from the effective reweighting efficiency.
Future analyses should use narrower realization-informed coefficient bounds, direct sampling from realization-based priors, or both.

\subsection{Application to an isolated point lens}
\label{sec:performance_PL}

\begin{figure*}[ht!]
    \includegraphics[width=0.9\textwidth]{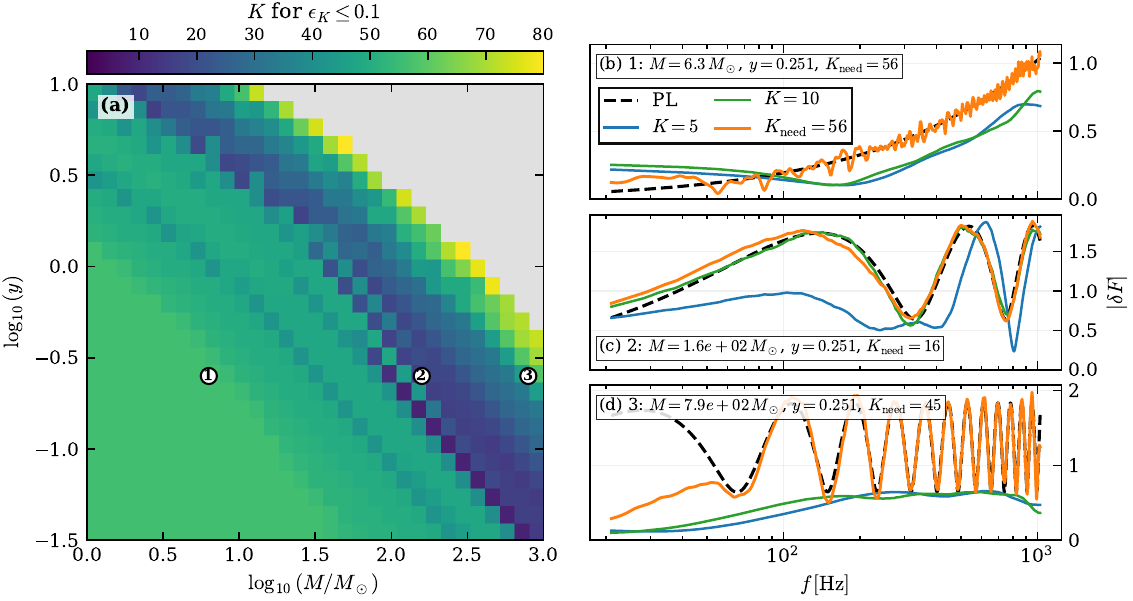}
    \caption{
    Reconstruction of isolated point-lens diffraction signals with the microlensing SVD basis.
    \textbf{Left:} Number of modes $K$ required to reach the weighted reconstruction error $\epsilon_K\leq 0.1$, Eq.~\eqref{eq:reconstruction_error}, as a function of lens mass $M$ and impact parameter $y$.
    Gray cells indicate configurations where the target accuracy is not reached with the retained basis.
    Three representative configurations are marked.
    \textbf{Right:} For these cases, the point-lens residual $|\delta F|$ is compared with reconstructions using different truncation orders.
    The basis captures the broad structure with relatively few modes, but rapid oscillatory features require many additional components.
    }
    \label{fig:pl_reconstruction}
\end{figure*}

We now apply the microlensing basis to the isolated point-lens model. 
This is a useful out-of-distribution test: an isolated point lens is deterministic and described by only a mass and an impact parameter, whereas the basis was trained on stochastic stellar fields embedded in an external macrolens. 
In the notation of Sec.~\ref{sec:lensing_stellar}, this limit corresponds to a single compact lens with no external convergence or shear. 
We compute the analytic point-lens amplification factor using \texttt{GLoW}~\cite{Villarrubia-Rojo:2024xcj}\footnote{\url{https://github.com/miguelzuma/GLoW_public}} and construct the residual $\delta F$ as in Eq.~\eqref{eq:delta_F}.

For each point-lens configuration $(M,y)$, where $M$ is the lens mass and $y$ the dimensionless impact parameter, we project the residual onto the microlensing basis and reconstruct it using the truncated expansion
\begin{equation}
    \delta F_K(f)
    =
    \sum_{k=1}^{K}
    \alpha_k \psi_k(f) .
\end{equation}
We quantify the reconstruction accuracy with the weighted fractional error
\begin{equation}
\epsilon_K^2
\equiv
\frac{
\left(\delta F-\delta F_K \mid \delta F-\delta F_K\right)_A
}{
\left(\delta F\mid\delta F\right)_A
}
=
1-
\frac{
\sum_{k=1}^{K}\alpha_k^2
}{
\left(\delta F\mid\delta F\right)_A
} ,
\label{eq:reconstruction_error}
\end{equation}
where the inner product is defined in Eq.~\eqref{eq:inner_product}. 
The last equality follows from the orthonormality of the basis with respect to the same weighted inner product.

Figure~\ref{fig:pl_reconstruction} shows the number of modes required to reach $\epsilon_K\leq0.1$ across the point-lens parameter space. 
Low-mass lenses and large impact parameters produce smoother residuals and can be described with fewer modes. 
More massive lenses, or lenses with smaller impact parameter, generate sharper interference features across the detector band and require many more modes. 
The gray region marks configurations for which the target accuracy is not reached with the retained basis.

The representative reconstructions in the right panels show the same behavior in frequency space. 
The first few modes reproduce the broad envelope of the point-lens residual, but many modes are needed to recover the rapidly oscillating phase and amplitude structure at high frequency. 
This inefficiency is expected. 
The basis was optimized to describe diffraction by stochastic stellar fields in a macrolens environment, not isolated deterministic point lenses. 
The point-lens test therefore shows that the SVD basis can represent out-of-distribution lensing signals in principle, but it also highlights a limitation: the basis does so much less efficiently than for the microlensing population on which it was trained.

\section{Conclusions and outlook} \label{sec:conclusions}

We have developed a reduced-order description of stochastic GW microlensing diffraction by stellar fields. 
Starting from numerical wave-optics amplification factors, we removed the smooth macroscopic contribution and constructed a basis for the residual diffraction pattern using a singular value decomposition (SVD). 
The resulting model represents microlensing through a manageable number of phenomenological coefficients $\{\alpha_k\}$, while retaining a direct connection to the simulated microlens population.

The construction separates two useful tasks. 
First, the microlensing coefficients can be sampled with broad priors as a flexible phenomenological model for frequency-dependent lensing distortions. 
Second, the same realization ensemble induces a coefficient distribution that can be used as a realization-based prior, or as a consistency check against stellar-field microlensing. 
This distinction is important: the basis provides flexibility, while the coefficient distribution supplies astrophysical regularization.

Our main results are:
\begin{itemize}
    \item \textit{A small number of modes captures most stellar-field diffraction.} 
    The first few microlensing modes retain most of the microlensing weight for the majority of realizations. 
    In particular, $K=8$ captures a median fraction $r_K\simeq0.95$ of the reference $K=80$ weight, with $q_{10}(r_K)\simeq0.83$; see Fig.~\ref{fig:truncation_weight}.

    \item \textit{The recovered coefficients carry information about the macrolens environment.} 
    The total microlensing weight $\upsilon_K$ correlates with stellar convergence, external shear, and macromagnification. 
    This suggests that measured microlensing coefficients can be used probabilistically to constrain the lens environment; see Figs.~\ref{fig:weights_summary_gamma_kappastar} and \ref{fig:upsilon_vs_mu}.

    \item \textit{Astrophysical priors sharpen inference and interpretation.} 
    In the simulated PE example, broad coefficient priors allow the microlensing model to absorb diffraction, but also introduce degeneracies with source parameters. 
    Reweighting by the realization-based prior reduces some of these degeneracies and produces microlensing parameters closer to the injected values; see Figs.~\ref{fig:source_weight_triangle} and \ref{fig:svd_coefficients_triangle}.

    \item \textit{The stellar-field basis is not an efficient representation of simple lens models.} 
    Although the basis can represent any modification of the waveform in principle, reproducing diffraction by an isolated point-lens requires many modes to capture the rapidly oscillatory features. 
    This confirms that the microlensing basis is specialized to the stochastic microlensing population on which it was trained; see Fig.~\ref{fig:pl_reconstruction}.
\end{itemize}
These conclusions should therefore be interpreted at two levels: the \modelclass{} construction and the use of broad coefficient priors are generic, while the quantitative mode hierarchy, realization-based priors, and correlations with $\gamma$, $\kappa_\star$, and $\mu_{\rm macro}$ are specific to the \modelname{} training ensemble.

Several extensions appear naturally. 
The basis can be tailored to specific sources by changing the frequency weight used to build the feature matrix: \modelname{} models adapted for low or high detector-frame masses will improve parameter estimation, while those adapted to simulated populations of strongly lensed signals will be better suited for matched-filter searches of microlensed GWs. 
This approach should reduce the number of coefficients required for targeted analyses, at the cost of making the model less generic.

Generalizations of the lens properties are also straightforward: these include variations on the macro-potential (macro-magnification, image parity) and the microlens population (convergence, mass function), which may include dark-matter objects~\cite{Zumalacarregui:2024ocb} in addition to stars and remnants.
these lens realizations will modify not only the microlensing basis, but also the associated parameter priors. An important extension is to connect the realization priors with astrophysical expectations: More realistic priors can be built by reweighting the simulation ensemble to simulated lens populations, and/or by using more flexible density estimators.

The coefficient distribution also opens a path toward astrophysical inference. 
Future work should develop estimators or likelihoods for macroscopic lens parameters, including $\mu_{\rm macro}$, $\kappa_\star$, image parity, and lens redshift, using either the recovered coefficient vector or compressed summaries such as $\upsilon_K$.

The model is also directly relevant for real-data searches. 
Matched-filter searches that include microlensing distortions may recover signals that would otherwise be missed or mischaracterized~\cite{Chan:2024qmb}. 
The microlensing model provides a practical way to incorporate such distortions without an explosion of parametric complexity. 
It could also be used to test whether candidate lensed events, such as GW231123\_135430 (hereafter GW231123)~\cite{GW231123_paper,Goyal:2025eqo,Shan:2025dcd,Chan:2025kyu}, are compatible with diffraction by a realistic stellar microlens population.

The same strategy can be generalized beyond the simple stellar fields and macro-potentials considered here. 
Important extensions include saddle-point macroimages and complex external potentials, such as caustics, where the source position has a strong impact on the wave-optics distortions~\cite{Lo:2024wqm,Ezquiaga:2025gkd,Vujeva:2025nwg}. 
Related basis constructions may also be useful for diffraction by compact dark matter substructure~\cite{Jung:2017flg,Choi:2021bkx,Urrutia:2021qak,Fairbairn:2022xln,GilChoi:2023qrz} --- whose ``dark timbre''~\cite{Urrutia:2024pos,Choi:2026lsa} is the imprint of a different lensing environment, analogous to the stellar timbre studied here --- and for space-based GW detectors, where lower frequencies probe different lens-mass scales~\cite{Gao:2021sxw,Savastano:2023spl,Caliskan:2023zqm,Brando:2024inp,Singh:2025uvp, Urrutia:2024pos,Ando:2026poq,Ando:2026eam,Choi:2026lsa}.
Ultimately, \modelname{} methods may become useful for modeling other GW sources subject to stochastic effects, such as random gas torques in accretion disks~\cite{Derdzinski:2020wlw,Zwick:2021dlg,Garg:2026jrv}, or microlensing of coherent electromagnetic sources, like fast radio bursts~\cite{Jow:2020rcy,CHIMEFRB:2022xzl,Leung:2022vcx,Tsai:2023tyw}.

\section*{Data availability}
The basis functions $\psi_k$ and the coefficients $\alpha_k$ of the 10,000 microlensing realizations used to build the \modelname{} model are available as Supplemental Material~\cite{SupplementalMaterial}, together with the detector weight $w(f)$, the macro-lens parameters of each realization and a stand-alone reader implementing the projection, reconstruction and realization priors described in Secs.~\ref{sec:modelbuild} and \ref{sec:properties}.
The simulated stellar-field realizations, the code that generates them and the code that constructs the reduced-order model will be made publicly available as part of a separate software release.

\acknowledgments{
\vspace{5pt}
\textit{Acknowledgements: We are very grateful to Alessandra Buonano, Mark Ho-Yeuk Cheung, Srashti Goyal, Hector Villarrubia-Rojo, Jay Wadekar and Matias Zaldarriaga for insightful discussions. 
MZ is grateful to the Institute for Advanced Study for hospitality during the completion of this work.
MZ acknowledges support from grant RYC2024-049805-I, funded by MICIU/AEI/10.13039/501100011033 and by the European Social Fund Plus (ESF+), from the Ayudas de Excelencia RYC-MaX 2024 programme of the Spanish National Research Council (CSIC), and from the Spanish Agencia Estatal de Investigación through grant “IFT Centro de Excelencia Severo Ochoa CEX2020-001007-S”. 
This work is supported by the ERC grants GLOW (101230608) and GWSky (101167314). Views and opinions expressed are however those of the author(s) only and do not necessarily reflect those of the European Union or the European Research Council Executive Agency. Neither the European Union nor the granting authority can be held responsible for them.
This material is based upon work supported by NSF's LIGO Laboratory which is a major facility fully funded by the National Science Foundation.
}
   }

\appendix

\bibliographystyle{apsrev4-2}
\bibliography{gw_lensing}

\end{document}